\documentclass{aa} 
\usepackage{ulem}
\usepackage{fullpage}
\usepackage{times}

\usepackage{appendix}
\usepackage[linesnumbered,ruled,vlined]{algorithm2e}
\usepackage{rotating}

\SetKwInput{KwInput}{Input}                
\SetKwInput{KwOutput}{Output}              
\DeclareMathOperator*{\argmin}{\arg\!\min}
\DeclareMathOperator*{\argmax}{\arg\!\max}
\newcommand{\norm}[1]{\left\lVert#1\right\rVert}
\newcommand{\DL}{$3\sigma$ detection limit }
\newcommand{\eSICL}{$e^{\rm{sep}}_{\rm{ICL}}$ }
\newcommand{\eCICL}{$e^{\rm{cont}}_{\rm{ICL}}$ }
\newcommand{\eSgal}{$e^{\rm{sep}}_{\rm{cluster}}$ }
\newcommand{\eCgal}{$e^{\rm{cont}}_{\rm{cluster}}$ }

\newcommand{\sbt}{$\mu_{\rm{SBT}} = 26.5$ mag.arcsec$^{-2}$}
\newcommand{\Ficl}{$f_{\rm{ICL}}$}
\newcommand{\Ficlsep}{$f^{\rm{sep}}_{\rm{ICL}}$}
\newcommand{\faint}{($z=0.5$, $r_{e}=150$~kpc and $f^{\rm{input}}_{\rm{ICL}} = 0.2$)}
\newcommand{\bright}{($r_{e}=50$~kpc and $f^{\rm{input}}_{\rm{ICL}} = 0.6$)}
\newcommand{\fluxiclsep}{$F^{\rm{sep}}_{\rm{ICL}}$}
\newcommand{\fluxicl}{$F_{\rm{ICL}}$}
\newcommand{\fluxgalsep}{$F^{\rm{sep}}_{\rm{cluster}}$}
\newcommand{\fluxgal}{$F_{\rm{cluster}}$}

\begin{document}

   \title{\texttt{DAWIS},}

   \subtitle{a Detection Algorithm with Wavelets for Intracluster light Studies}

    \titlerunning{Presenting \texttt{DAWIS}}

   \author{A. Ellien\inst{1}
        \and
           E. Slezak\inst{2}
        \and
           N. Martinet\inst{3}
        \and
           F. Durret\inst{1}
        \and 
            C. Adami\inst{3}
        \and 
            R. Gavazzi\inst{1}
        \and
           C. R. Raba\c ca\inst{4}
        \and
           C. Da Rocha\inst{5}
         \and
          D. N. Epit\'acio Pereira\inst{5}
        }

   \institute{Sorbonne Universit\'e, CNRS, UMR 7095, Institut d'Astrophysique de Paris, 98bis Bd Arago, F-75014 Paris\\
            \email{ellien@iap.fr}
        \and
            Observatoire de la C\^{o}te d'Azur, BP 4229, F-06304 Nice cedex4, France
        \and
            Aix-Marseille Univ., CNRS, CNES, LAM, Marseille, France
         \and 
            UFRJ, Observatório do Valongo, Rio de Janeiro, RJ, Brazi
        \and
            Independent Researcher
          }
   \date{Accepted January 11, 2021}

 
  \abstract
   {Large amounts of deep optical images will be available in the near future, allowing statistically significant studies of low surface brightness structures such as intracluster light (ICL) in galaxy clusters. The detection of these structures requires efficient algorithms dedicated to this task, where traditional methods suffer difficulties.
   }
   {We present our new Detection Algorithm with Wavelets for Intracluster light Studies (\texttt{DAWIS}), developed and optimised for the detection of low surface brightness sources in images, in particular (but not limited to) ICL.
   }
   {\texttt{DAWIS} follows a multiresolution vision based on wavelet representation to detect sources, embedded in an iterative procedure called synthesis-by-analysis approach to restore the complete unmasked light distribution of these sources with very good quality. The algorithm is built so sources can be classified based on criteria depending on the analysis goal; we display in this work the case of ICL detection and the measurement of ICL fractions. We test the efficiency of \texttt{DAWIS} on 270 mock images of galaxy clusters with various ICL profiles and compare its efficiency to more traditional ICL detection methods such as the surface brightness threshold method. We also run \texttt{DAWIS} on a real galaxy cluster image, and compare the output to results obtained with previous multiscale analysis algorithms.
   }
   {We find in simulations that in average \texttt{DAWIS} is able to disentangle galaxy light from ICL more efficiently, and to detect a greater quantity of ICL flux due to the way it handles sky background noise. We also show that the ICL fraction, a metric used on a regular basis to characterise ICL, is subject to several measurement biases both on galaxies and ICL fluxes. In the real galaxy cluster image, \texttt{DAWIS} detects a faint and extended source with an absolute magnitude two orders brighter than previous multiscale methods. 
   }
   {}


   \maketitle
%

\section{Introduction}

Low surface brightness (LSB) science will improve this new decade with the launch of several large observational programs. The \textit{Vera Rubin Observatory Large Synoptic Survey Telescope} \citep[LSST;][]{Ivezic2019}, a ground-based system featuring an 8.4 meter primary mirror, will lead a ten year survey on a 18,000~deg$^{2}$ sky area, reaching a foreseen limiting depth of $\mu_g = 31$~mag~arcsec$^{-2}$. In space, the \textit{Euclid} mission will perform three deep field programs in the VIS broad band (R+I+Z) covering in total 40 deg$^{2}$ with a conservatively estimated limiting magnitude of $\mu_{\rm{VIS}} = 26.5$~mag~arcsec$^{-2}$. New missions such as the MESSIER surveyor \citep{Valls-Gabaud2017}, a space telescope optimized specifically for LSB imaging both in the UV and the visible wavelengths, are also planned for the upcoming years.

These new programs will complement current and past LSB surveys (of which we only give a non exhaustive review here). Small telescopes optimized for LSB imaging such as the \textit{Dragonfly Telephoto Array} \citep{Abraham2014} or the \textit{Burrell Schmidt Telescope} \citep{Mihos2017} are obtaining good results from the ground, reaching limiting depths of $\mu_{g}=29.5$~mag~arcsec$^{-2}$. These small telescopes take advantage of the minimization of artificial contamination sources, in a field where other instruments not originally dedicated to this kind of studies find strong disadvantages in that regard, such as the \textit{MegaCam} instrument on the \textit{Canada France Hawaii Telescope} (CFHT). This last instrument has nevertheless achieved its share of surveys as its limiting depth has been pushed by constraining instrumental contamination effects through refined observational strategies and reduction softwares. The CFHT \textit{Legacy Survey} \citep{Gwyn2012} is a good example, as it allowed to detect LSB structures such as tidal streams \citep{Atkinson2013}, followed later by the \textit{Next Generation Virgo Cluster Survey} \citep{Ferrarese2012}, a survey dedicated to deep imaging of the Virgo cluster. Next in line is the ongoing \textit{Ultraviolet Near-Infrared Optical Northern Survey} \citep[UNIONS;][]{Ibata2017}, which features images processed with the \texttt{Elixir-LSB} pipeline \citep{Duc2011} and reaching a limiting depth of $\mu_{r}=28.3$ mag~arcsec$^{-2}$ in the $r$-band on a wide sky area of $\sim$~5000 deg$^2$.

As these ground-based instruments are still limited by the atmosphere, surveys to capture LSB features have also been led from space with the \textit{Hubble Space Telescope} (HST), the prime example being the \textit{HST Ultra Deep Field} \citep{Beckwith2006}. However, the small field-of-view (FoV) of the HST does not allow to probe large spatial extents, leading to different scientific goals such as the study of distant objects which are smaller on the projected sky plane (typically high redshift galaxies or galaxy clusters). The ongoing \textit{Beyond Ultra-deep Frontier Field And Legacy Observations} survey \citep[BUFFALO;][]{Steinhart2020}, next in line to the \textit{Hubble Frontier Field} \citep[HFF;][]{Lotz2017}, follows this trend and targets with very deep imaging six massive galaxy clusters with redshifts in the range $0.3 < z < 0.6$.


Compared to the ongoing surveys which are all limited in their own ways, the new generation of telescopes will bring an unprecedented amount of data to exploit. As a sub-field of LSB science, the detection and analysis of intracluster light (ICL) will strongly be impacted, as one of the primary requirements to study this faint feature of galaxy clusters in the visible bands is the gathering of deep images. However, the state of research in this field is presently not well defined, as there is no consensus either on a strict definition of ICL in astronomical images or on the best method to detect it. This leads to a variety of studies on this subject which are barely comparable in view of the large discrepancies implied by the methodologies used \citep[for more details, see the review by ][]{Montes2019b}. The same can be said about surface brightness depth and the methodology to compute detection limits from the sky background in images as explained by \citet{Mihos2019}. Before tackling the large upcoming data sets, a much needed analysis on the currently used detection method properties should be led.

As part of this new challenge in LSB astronomy is purely technical (processing great numbers of images is often CPU-time expensive), new algorithms with an emphasis on efficiency need to be developed to analyse images and to capture the useful information they contain. This is an ideal time to (re)explore concepts from signal and image analysis and adapt them to LSB astronomy. With this in mind we developed \texttt{DAWIS}, an algorithm optimised for the detection of LSB sources and highly parallelized to be run on large samples of images. 

Such a new algorithm needs to be tested on simulations and contrasted with previous detection methods. To run the tests, we create images of simulated galaxy clusters and ICL using the \texttt{Galsim} package \citep{Rowe2015}. These images only reproduce the photometric aspect of galaxy clusters and cannot be used to draw conclusions on the properties of ICL. However their content is a known value, which allowed us to estimate the efficiency of \texttt{DAWIS} and of previous methods for detecting ICL. We were also able to constrain the different biases and contamination effects occurring when applying these methods.

This paper is organised as follows. In Section~\ref{sec:intracluster_light_detection} we give some context on the various detection methods used to detect ICL in deep images and different effects limiting and contaminating this detection. In Section~\ref{sec:dawis} we present the technicalities of \texttt{DAWIS} and the core of the algorithm. In Section~\ref{sec:simulations} we describe how we simulated mock images of galaxy clusters with ICL with the modeling package \texttt{Galsim}. In Section~\ref{sec:application_of_detection_methods} we apply the different detection methods to the simulations and present the results. In Section~\ref{sec:illustration_on_real_data}, we run \texttt{DAWIS} on real data, and compare the results with previous works. In Section~\ref{sec:conclusions} we discuss the results and the performances of the methods, both on simulations and on real data. We assume a standard $\Lambda$CDM cosmology with $\Omega_{m}=0.3$, $\Omega_{\Lambda}=0.7$ and $H_{0}=70$~km.s$^{-1}$.Mpc$^{-1}$.


\section{Overview of ICL detection methods}
\label{sec:intracluster_light_detection}

Behind the term \textit{intracluster light detection} are hiding several choices, starting with the way images are acquired (space against ground, survey strategies with or without image dithering, long exposures or stacks of short ones), data reduction (depending on the properties of the instrument), and finally how the useful information is disentangled from noise and contamination sources (sky background estimation, separation of ICL from galaxies and foreground objects, point spread function wings, scattered light). We do not try to address here all these effects at once but tackle the most fundamental aspect of ICL detection: disentangling the ICL component from galaxy luminosity distributions and taking into account the sky background noise.

A large variety of methods have been used to disentangle ICL from bright sources in astronomical images. We group them roughly into three categories: the Surface Brightness Threshold (SBT) methods, the Profile Fitting (PF) methods, and the multiscale analysis (usually making use of wavelet bases) methods. Note that there are in the literature as many procedures to detect the ICL as there are papers studying it. This classification is therefore simplifying the larger picture, but is probably a good start to try to characterize and evaluate this great variety of approaches and ICL definitions.

\subsection{The Surface Brightness Threshold methods}

The first method is the SBT method, which consists in applying a predefined surface brightness threshold in the image, in order to demarcate the ICL from the galaxy luminosity profiles. The most common threshold value is given by the Holmberg radius \citep[defined by the isophote $\mu_{B}=26.5$ mag~arcsec$^{-2}$;][]{Holmberg1958}, which delimits at first order the geometric size of a galaxy in optical images. This threshold has been used in different cases, either to mask the cluster's galaxies (the threshold defines in that case the limit of the galaxy extension), or simply to separate the outskirts of the Brightest Cluster Galaxy (BCG) luminosity profile from the ICL. In any case, the threshold acts as a decision operator, attributing pixels either to galaxies or to ICL.

This method is quick to implement and has been used in several works \citep[][just to cite a few]{Krick2006, Krick2007, Burke2012, DeMaio2018, Ko2018, Montes2018}. Different values for the threshold have been tested \citep{Feldmeier2002, Feldmeier2004}, resulting in large discrepancies for the results obtained in observational data \citep{Kluge2020}. From the $N$-body and hydrodynamical simulations side, studies have shown similar results \citep{Rudick2011, Tang2018}. Depending on the choice of the person doing the study, different SBT values have been tested with values ranging from $\mu_{\rm{SBT}}\sim23$ to $\mu_{\rm{SBT}}\sim27$ mag~arcsec$^{-2}$ in the V band, showcasing once again high discrepancies between them. However, due to the difficulty to define ICL even in simulations, authors proposed that different definitions would naturally lead to very different results, without specifying which one is the most appropriate.

\subsection{The Profile Fitting methods}

Intensity profile fitting (PF) methods consist in fitting analytical functions to the intensity distribution of the galaxies or to the ICL. The first use of this approach was made by fitting de Vaucouleurs (dV) profiles to the inner intensity distribution of BCGs, before detecting excesses of light in the outskirts which were attributed to ICL \citep{Uson1991, Scheik1994, Feldmeier2002, Zibetti2005}. Since the BCG and the ICL intensity distributions blend together smoothly, several studies tried to characterise the BCG+ICL intensity distribution, using a single S\'ersic profile \citep{Krick2006, Krick2007}. Other authors tried to distinguish both distributions by fitting sums of profiles, such as double S\'ersic \citep{Seigar2007, Durret2019, Kluge2020}, double exponential profiles \citep{Gonzalez2005} or composite profiles \citep{Donzelli2011}. In some cases different analytical functions can correctly fit the same distribution \citep{Puchwein2010}, which makes the physical interpretation of such results complicated. Most notably, recent works have qualified the decomposition of the BCG+ICL distribution into separate luminosity profiles as likely unphysical \citep{Remus2017, Kluge2020}.

Fitting algorithms such as \texttt{Galfit} usually decompose galaxy intensity profiles into two components (bulge plus disk), and allow to model the radial distribution by S\'ersic profiles, while the angular distribution is controlled by trigonometric functions. This allows to fit a great diversity of intensity profiles, as long as it does not present too sharp features. Most of galaxy intensity profiles can therefore be fitted by such methods. But for complex objects such as strongly interacting galaxies, a high level of interactivity is required, as the user needs in these cases to adjust manually the parameters involved in the fitting procedure. This makes this method difficult to fully automatize, which is a downside when many galaxies are present in the FoV \citep[but not impossible as shown in][]{Morishita2017}. Additionally, the blending of galaxy intensity profiles in the high density regions of galaxy clusters is also another problem for this kind of approach.

Recently, more sophisticated fitting algorithms were developed, most notably \texttt{CICLE} \citep{Jimenez-Teja2012, Jimenez-Teja2016}. \texttt{CICLE} models galaxy luminosity profiles with Chebyshev Fourier (CHEFs) functions. These forms are implemented into a fitting pipeline using outputs (position and size of the object) from the \texttt{SExtractor} image analysis software \citep{Bertin1996}, as a subjective origin needs to be set for the basis function when modelling each galaxy. Therefore, while this fitting method is very strong to accurately model the surface brightness distribution of detected objects, it is still sensitive to the detection performances of \texttt{SExtractor} or other detection methods used beforehand.

\subsection{Wavelets and multiscale image analysis methods}

Another approach to the detection of ICL (and generically of faint and extended sources among much brighter objects in astronomical images) has been the use of multiscale, wavelet-based algorithms. Isotropic wavelet bases such as the B3-spline scaling function and its associated wavelet transform has been used for the first time in an astronomical context by \citet[][]{Slezak1994} to detect the large intracluster medium halo in X-ray images of galaxy clusters. The next year, \citet[][]{Bijaoui1995} devised a powerful multiresolution vision model to analyse the three dimensional (3D) data set of wavelet coefficients generated by an isotropic wavelet transform. The related procedure allows to detect significant structures in the 2D wavelet domains, to identify objects in this 3D wavelet space, and to restore the denoised luminosity distribution of these detected structures. Such an approach endorses the detection of extended sources, and can therefore be adapted to the detection of ICL in optical images of galaxy clusters, as demonstrated by \citet[][]{Adami2005}. Another known implementation has been the \texttt{OV\_WAV} package, developed in 2003 at Observatório do Valongo (OV-UFRJ, Rio de Janeiro) by Daniel Epitácio Pereira and Carlos Rabaça. This \texttt{IDL} package has been used in several works to detect diffuse light in galaxy groups or clusters \citep{DaRocha2005, DaRocha2008, Guennou2012, Adami2013}. More recently, another implementation of the multiresolution vision model has been developed for processing astronomical MHz radio images, adding an extra deconvolution step and describing the problem under the modern framework of sparse representation \citep[the \texttt{MORESANE} radio astronomical image analysis algorithm; ][]{Dabbech2015}. Hereafter we give an in depth description of this kind of approach and its mathematical background, compiling and unifying the various informations contained in previous articles.

\subsubsection{Choosing a suitable representation space}

Beyond the applied algorithm itself, the efficiency of such an approach is tightly connected to the mathematical space used to represent the information content of the signal. One way to carry out the detection of the ICL is then to find a new space which highlights the low surface brightness and large spatial extent of the ICL, easing its discrimination from other astronomical sources.

The mathematical space where the signal is initially represented (hereafter the direct space) may not be the most efficient one for the pursued goal of our our analysis. This can specially be the case when the information of interest is strongly mixed with other (in that case) components. So, identifying a new representation space (and consequently the set of basis functions generating this space) for the data is then of uttermost importance for the final result of the processing. The transform of the initial signal from the direct space to this new space is obtained through inner products with the set of basis functions, defining what is called a projection. Since detecting the ICL in direct space is a difficult task with traditional methods, one can search for a more suitable representation space to make it easier. 

There is a large choice of basis functions with different regularity properties, which can define a basis, orthogonal or not, and consequently of representation spaces. The choice of which basis to use then depends on the signal characteristics to be strengthened according to the analysis goals. A generic approach for this choice is the notion of sparsity: an adequate function basis disentangles the signal of interest from the rest by concentrating the useful information on few high-valued coefficients while spreading the noise and worthless components on many coefficients with low values. Note that while a sparse representation gathers the relevant information, it may simplify it, as features very different from the basis functions are lost.

In our case, the typical image of a galaxy cluster in visible bands can roughly be hierarchically decomposed into several circular or elliptical components with different characteristic sizes and intensities: the brightest sources in the field-of-view are usually PSF-like Milky Way stars and foreground galaxies, then galaxies at the cluster redshift and finally faint more distant galaxies, and even fainter extended sources such as the ICL, all of them being superimposed on a spatially slowly varying (instrumental or not) sky background.

Due to their elliptical shapes when projected onto the sky, a relevant first approach to describe these sources is to use isotropic functions which are azimuthally invariant. Since these sources may have very different characteristic sizes (or spatial frequencies), the function basis also needs to capture localized information (high frequencies) as well as mean behaviours related to the information averaged over a given region of the image (low frequencies). The  transform associated to  such a basis is  called a multiscale transform. It enables one to study  the signal at varying scales: the  transform at small scales gives access to the thinnest and local features of the signal while its transform at large scales captures its overall behavior. As the analysis goal here is the detection of any feature  of interest, one would also prefer to rely on an analysis which is invariant under translations. The inner products of  the signal with the basis functions then do not depend on the position of the information in the signal, meaning that there is no need to set a subjective origin for the transform, as it is the case for shapelets \citep{Refregier2003}. 

In our case, an interesting class of functions is the wavelet family. The first (admissibility condition) and second order moments of wavelets are equal to zero, making them contrast detectors in the simplest form. An example of such a function is the Morlet wavelet which is basically a cosine weighted by a Gaussian. A wavelet basis is built by shifting and dilating the same wavelet function (the so called mother wavelet), and the associated transform is called a wavelet transform. Contrary to the Fourier basis, which gives the most accurate frequency information in an infinite temporal signal in exchange of the loss of date information, wavelet transforms are part of the multiscale transform family. They therefore allow a time-frequency representation, as different frequency scales are locally explored by different dilated and translated versions of the same mother wavelet function, providing thereby a date information for each analyzed sample from the whole data set.

The 2D isotropic wavelet bases satisfy the criteria listed above and are consequently adequate at first order to study astronomical images, specifically images with both bright localized sources (the signature in the image of objects like stars or galaxies) and large diffuse sources (the signature of objects like intracluster light halos). On the other hand, 2D isotropic wavelet bases are certainly not suitable to detect in the image features which are strongly anisotropic or with very sharp edges, such as elongated rectangles or lines (one could think of the point spread function spikes around stars, cosmic rays, satellite trails or tidal debris around galaxies for example). Indeed, any such transform acts as a measure of similarity between the set of wavelets and the features under scrutiny. Making use of an isotropic filter implies loose information related to anisotropies if any. A popular example of a 2D isotropic wavelet function is the normalized second derivative of the Gaussian, nicknamed the Mexican Hat (MH) due its shape of a sombrero, when used as a 2D kernel : a disk of positive values surrounded by an annulus of negative ones, the integral of which is normalized to zero.

\subsubsection{Discretization and multiresolution approach}

In order to use wavelet transforms to analyze images, one needs to implement them into algorithms and to discretize the continuous theoretical functions along both  scales and  image axes. A discrete set of functions is built which may constitute what is called a frame according to the chosen discretization scheme for wavelets with "good" enough regularity properties. This discretization is not without consequence, as part of the information contained in the continuous function might be lost. An upper limit for the  loss of information for a given set of frame bounds, i.e. a discretization scheme, can be computed \citep{Daubechies1990}. This loss is small for the MH function when one considers a dyadic scheme and two voices per octave\footnote{The scale parameter has to be discretized when considering a discrete wavelet transform. This is usually done by raising an initial scale to positive $j$ integer powers. A dyadic scheme involves powers of 2, so that the different scales are obtained using a factor $2^{j/\nu}$ where $\nu$ is an integer parameter greater than one often referred to as the number of "voices per octave".}.

A problem with the MH function though is its extended spatial support - the fact that its profile extends to infinity. It is in practice numerically impossible to compute the exact theoretical transform as approximations need to be done at the edges. Consequently, a widely used MH-like function is the B3-spline wavelet, which is also isotropic and translation invariant with a controlled loss of information when using dyadic scales, but with the prime advantage of having a compact support making the transform computable without any approximation.

A major breakthrough for the understanding and efficient implementation of wavelets into algorithms resulted from the multiresolution theory of \citet[][]{Mallat1989}, showing that the set of wavelet functions are no more than a hierarchy or cascades of filters, also known as filter banks in the signal processing application domain. Within this framework, the mother wavelets are defined by means of a scaling function, which acts as a low-pass filter. This mother wavelet function appears to be in fact the difference between this scaling function and a normalized version of it dilated by a factor 2 in size. For instance, the MH is the difference between two differently scaled Gaussian functions, and the B3-spline wavelet is the difference between two differently scaled B3-spline functions. 

The link between some classes of wavelets (e.g. B3-spline) and filter banks is also expressed through a dilation equation: the scaling function at scale 2 can be expressed as a linear combination of these scaling functions at scale 1. This is true for the continuous basis as well as for the dyadically discretized version of it. Therefore, one can convolve iteratively an image with dilated (and decimated) B3-spline functions using a dyadic scheme, building in this way a set of coarser and coarser approximations of the initial 2D signal. The difference between two successive levels of approximation then gives the  wavelet coefficients related to this scale range. These coefficients can be viewed as a measure of the information difference between the coarser approximation and the thinner one, or in other words of the details in the image with typical sizes within these two scales (cf. band-pass filter).

This iterative approach, which makes use of the so-called \textit{à trous} algorithm from \citet[][cf. spatial decimation of the low-pass filters]{Holschneider1989}, is much faster than using convolutions with filters of increasing supports to compute the transform. It does not rely on numeric integrals but benefits from a simplified filtering operator based on simple multiplications and additions. There are different versions of this algorithm depending on the analysis goal. For the 2D decimated wavelet transform, the size in pixels of each smoothed image is divided by four with respect to the previous level of approximation (and so are the number of associated wavelet coefficients). This leads to a pyramidal representation that is well suited to encode at a given level the features of the image with different sizes in a sparse way when there is a good match between these features and the scaling functions. An undecimated version has also been proposed, which allows to keep precise spatial information since all the wavelet planes have the same size as the original image. This undecimated wavelet transform with the \textit{à trous} algorithm and the B3-spline scaling function basis is central to the multiresolution vision model of \citet{Bijaoui1995}, a basic conceptual framework for denoising or source detection algorithms.

\subsubsection{Analysis and Restoration}
\label{subsec:interscale_connectivity}
Besides the Haar wavelet, Daubechies has proved \citep[see for instance][]{Daubechies1992} that a wavelet basis cannot have at the same time a compact support, be isotropic and be orthogonal. Since the B3-spline wavelet basis has a compact support and is isotropic, its associated representation space is not  orthogonal: a source with a single characteristic size in the image will then not be seen as a set of wavelet coefficients with high values at one single scale, but will have non-null wavelet coefficients at several successive scales. An analysis of the wavelet coefficients is then needed along the spatial axes and the scale axis to link together these coefficients and correctly characterize the associated source in the wavelet domain. For the spatial analysis, the undecimated \textit{à trous} algorithm of \citet[][]{Holschneider1989}, first used in the astronomical context in  \citet[][]{Slezak1990}, and also advertised as the Isotropic Undecimated Wavelet Transform \citep[IUWT,][]{Starck2007}, is easier to use since the various wavelet planes have the same size as the original image.

We explain briefly hereafter the properties of the IUWT, and a rigorous definition is given in Section~\ref{subsec:IUWT}. The image is decomposed according to a dyadic scheme into several scales that exhibit sources with the same characteristic size: the first few high frequency scales contain  compact sources (small scale details), while the low frequency scales contain extended sources (large scale details). Though not perfect since objects in the original image are spread through several sources at different scales, the IUWT is a sparse representation of the initial data. Providing that the noise affecting the data is white, objects in the direct space generate indeed wavelet coefficients with much higher values than those related to the noise-dominated pixels at any scale except the smallest one.

The fact that the WT is sparse makes the detection of any faint but extended source much easier in the wavelet domain than in the direct space, especially at the large scales relevant for the ICL component. A hard thresholding of the wavelet coefficients is therefore an efficient way to denoise the data and for instance detect objects or structures. To do so, one only needs to select scale by scale the significant wavelet coefficients and group them into connected domains (see Section~\ref{subsec:filtering_support}). Restoring an image in the direct space for a single detected object is a little more tricky: an interscale analysis must be performed to build interscale trees using the spatial and scale positions of each significant domain, and various constraints can be applied when building or pruning these trees. The information from a pruned interscale tree can then be used to restore (or reconstruct) the associated object intensity distribution in the direct space.

Following the method described in \citet[][]{Bijaoui1995}, only the wavelet coefficients of the domain where the interscale maximum of a tree is located (i.e. the region with the highest value within the tree, hence with the highest information content), and from every region linked to it at smaller scales, are used. This pruning of the interscale trees ensures that the restoration algorithm has access to enough information to compute a satisfying solution, and that the information retained does belong to the same structure in the direct image. However, such a pruning discards the information from domains at lower spatial frequencies than the wavelet scale of the interscale maximum. Hence, in case of a source with an intensity profile featuring for instance an inner core much brighter than its outskirts, only the bright core will be reconstructed while most of the outskirts will be missed. Faint sources near bright ones will therefore be correctly processed only if an analysis with such a pruning is performed at least twice.

The restoration step is applied to each tree individually and can be viewed as an inverse problem that yields to an iterative estimation process (see Section~\ref{subsec:inverse_problem}). Several solutions to such optimization problems have been proposed in the literature such as conjugate gradient methods or the Landweber scheme \citep{Starck1998}, based on positivity and other regularization constraints. In a more straightforward way these estimation algorithms aim to reproduce the direct space intensity profile of the detected object by adding and subtracting different elements from a wavelet basis, and using information from the interscale tree. In this paper, the wavelet basis used for the restoration step is usually the same as the one used for the analysis (the B3-spline wavelet basis).

There are several remarks to make on this overall method. The first one is that it is parameter prior-free, as there is no need to specify a profile for the objects that are reconstructed contrary to usual fitting methods. Keep in mind though that a choice is made through the wavelet basis used for the analysis and the restoration. 

Astronomical sources cannot be represented by a single wavelet function, but rather by linear combinations of elements from the same wavelet basis. This implies a selection (made through the estimation algorithm) of the elements of the basis function leading to the best representation. Such a selection is performed to minimize the difference in shape between the source intensity distribution and the pattern in the direct space linked to the set of wavelet functions used to model it. However, this selection is almost always sub-optimal, and will generically result in artifacts in the restored profile. In the best case scenario, the amplitude of these artifacts is very low compared to the other source distribution attributes. But sometimes, the iterative process fails to compensate for this difference (and may even amplify it in the worst case scenario of strongly overlapping objects with high surface brightness), and artifacts can then be significant. In our case, due to the nature of the wavelet pattern, which is a disk of positive pixels surrounded by an annulus of negative ones, these artifacts if any take the form of spurious rings around restored sources. Likewise, choosing an isotropic vision model also leads to slight morphological biases on the reconstruction of anisotropic objects, typically  galaxies with large ellipticities for which the solution has the same integrated flux but with a more circular light profile.

\subsubsection{Implementation and limitations}
\label{subsec:implementation_and_limitations}
Such a wavelet-based multiscale approach has been first used by Adami et al. (2005) to detect a large scale diffuse component within the Coma cluster of galaxies. The \texttt{IDL} package \texttt{OV\_WAV} is another implementation of this multiresolution approach, and was used to detect diffuse sources in astronomical images of galaxy groups \citep{DaRocha2005}. The wavelet representation allowed them to detect extended sources down to $(S/N)\sim0.1$ per pixel, enough to characterize the intragroup light (IGL) of HCG 15, HCG~35 and HCG~51 \citep{DaRocha2008}.


Even with fast \textit{à trous} algorithms, such an analysis procedure is computationally time expensive, mainly due to the interscale analysis and to the object reconstructions. In addition, a problem met by this approach is the false detections due to statistical fluctuations of the noise. As previously mentioned, the noise is dominant in the high frequency wavelet scales, and packs of noise pixels with high values can be detected as sources and have their own interscale tree. This results in the reconstruction of wrong detections situated in the high-frequency scales, increasing the computing time. The authors of \texttt{OV\_WAV} used various ways of thresholding their wavelet scales in order to limit the false detections, applying higher thresholds for the high frequency scales, without completely solving the problem.

Running \texttt{OV\_WAV} on an image allows to detect most of the bright sources and to reconstruct them properly up to a very high precision. All the reconstructed sources are then concatenated into a single image: the full reconstructed image of the original field. A residual image can then be computed by subtracting the reconstructed image to the original one.

Due to various reconstruction factors described in \ref{subsec:interscale_connectivity}, low surface brightness features can be missed. \citet[][]{Adami2005} had the idea of running the algorithm a second time, but on the residual image, in order to detect outer halos of galaxies and other more diffuse structures. While better results are obtained in this way, one must note that the overall performance of this iterative approach is still determined by the intrinsic quality of the restored intensity distribution for the detected source. It is especially the case for very peaked and bright sources, as any high value residual left from the first pass could then be detected as a significant structure in the second one, once again hiding faint sources superimposed or close to it. \citet{Ellien2019} chose this approach for a beta version of \texttt{DAWIS}, where the same algorithm was run three times in a row to correctly detect and model every galaxy in the image. After such a procedure, if the ICL is not detected with the wavelet algorithm, it is possible as a fast alternative to detect it in the final residual image by applying an appropriate standard sky background threshold. In that case, the wavelet analysis acted as a simple modeling tool for galaxies, analogous to profile fitting methods. One could note that a fully iterative procedure with this kind of wavelet algorithm is always difficult to apply, due to its computational cost, but seems to be the best way to significantly improve the overall quality of the analysis (especially with regard to object restoration), and to thoroughly detect ICL in the wavelet space.

In parallel to the detection of ICL, this multiscale approach has been adapted to different kinds of data, where the scientific goals are similar (e.g., detecting faint and extended sources hidden by bright and compact sources). Most notably, \citet{Dabbech2015} proposed the algorithm \texttt{MORESANE} (\textit{MOdel REconstruction by Synthesis-ANalysis Estimators}), developed for processing radio images of galaxy clusters taking into account the complex PSF of radio interferometers. As already said, this algorithm makes use of the multiresolution vision model of \citet{Bijaoui1995}, embedded in an iterative procedure generalizing and upgrading the process implied by the earlier works of \citet{Adami2005} on ICL. This allows to solve most of the problems described in the previous paragraphs. The author of \citet{Dabbech2015} also provided a description of the overall procedure and algorithm in terms of sparse representation,
called \textit{synthesis-by-analysis approach}. We decided to use this latest version of multiscale image analysis procedure as a starting point to upgrade this class of methods used to detect ICL. We propose here our own version of this strategy, optimized both for computation time and for optical images, which is presented in the next section.


\section{\texttt{DAWIS}}
\label{sec:dawis}
In this section we present the operating structure of \texttt{DAWIS}. While many notions addressed here are already well known, we detail them anyway with in mind the global understanding of the algorithm. Readers solely interested in the ICL detection performance method comparison can jump to Section~\ref{sec:simulations}.
We use the following notations: matrices are denoted by bold upper case letters (e.g., $\mathbf{A}$ with a transpose $\mathbf{A}\!^\top$), vectors by bold lower case letters (e.g., $\mathbf{v}$). A component of row index i and column index j is given by $\mathbf{A}_{i,j}$. A vector component of index i is given by $\mathbf{v}_{i}$. Vectors are all column vectors, and row vectors are denoted as transposes of column vectors (e.g., $\mathbf{v}\!^{\top}$). Vector subsets and matrix columns or rows are denoted by top or bottom indexes with parentheses (e.g., $\mathbf{v}^{(j)}$ or $\mathbf{v}_{(j)}$).

\subsection{Inverse problem and sparse representations}
\label{subsec:inverse_problem}

When modelling a signal from observed data, the solution of this inverse problem may not be unique. To solve such a so-called ill-posed problem, one must introduce a penalty term in the mathematical equation describing it to select a particular solution. This solution must satisfy this added criterion, thereby leading to an optimization problem. Let us consider the generic equation:
\begin{equation}
    \mathbf{y} = \mathbf{\mathcal{H} x + n},
\end{equation}
\noindent
where $\mathbf{y} \in \mathbb{R}^{M}$ is the measured signal, $\mathbf{x} \in \mathbb{R}^{N}$ is the initial signal, $\mathbf{\mathcal{H}}:\mathbb{R}^{N}\to\mathbb{R}^{M}$ is a known (or approximately known) degradation operator, and $\mathbf{n} \in \mathbb{R}^{M}$ is an additive noise. This structure can be used to represent many problems in image and signal processing such as denoising with $\mathcal{H}=\mathbf{I}$ or deconvolution with $\mathcal{H}$ an impulse response (i.e. the PSF for a focused optical system). Recovering the initial signal from the observed (sub-)set of data is an inverse problem which can be solved via a penalized estimation process written as :

\begin{equation}
    \label{eq:general_solution_inverse_problem}
    \hat{\mathbf{x}} = \argmin_{\mathbf{x'}} \frac{1}{2} \norm{\mathbf{
y} - \mathbf{\mathcal{H} x'}}^{2}_{2} + \lambda \mathcal{R}(\mathbf{x'}),
\end{equation}

\noindent
where $\mathcal{R}:\mathbb{R}^{N}\to\mathbb{R}^{+}$ is the penalization function and $\lambda\in\mathbb{R}^{+}$ is the regularization parameter. One widely used constraint is the Tikhonov regularization which may rely for instance on difference operators to promote a smooth solution or the identity matrix to give preference to solutions with small $\ell_{p}$ norms\footnote{The $\ell_{p}$ norm is given by $\ell^{p}_{p}=\sum_{i} \left|\mathbf{x}_{i}\right|^{p}$ for a vector $\mathbf{x}$.}. One can also decide to enforce sparsity of the solution for a given representation space. To do so, the penalized function $\mathcal{R}$ is then a measure $\mathcal{S}$ of the sparsity of the solution when projected onto the basis defining this new space, i.e. after having applied to this solution a transform $\gamma$, so that the penalty term is written as $\mathcal{R}(\mathbf{x'})=\mathcal{S}(\boldsymbol{\gamma})$ with $\boldsymbol{\gamma}=\gamma(\mathbf{x'})$. This transform $\gamma$ is usually chosen to be a linear operator. The normalized matrix aggregating the new basis functions as columns is commonly referred to as a dictionary and each column or vector of it is then an element also called an atom. 

A natural choice for the function $\mathcal{S}$ is the $\ell_{p}$ norm with $0<p<1$ to favour sparsity. Case $p=0$ related to support minimization is usually untractable since highly nonconvex, hence a NP-hard problem; case $p=1$ corresponds to the tightest convex relaxation to this problem, which may still not be easy to solve efficiently when the dimension is high. The use of dictionaries (cf. composite features) in combination with $\ell_{1}$ norm gave rise to several minimization algorithms with many variants to find the best approximation of $\mathbf{x}$ by the elements of the dictionary, such as the Method of Frame \citep{Daubechies1988}, the Basis Pursuit scheme \citep{Chen2001} or the Compressive Sensing \citep{Donoho2006}. Faster than convex optimization but lacking uniformity, a greedy method such as (Orthogonal) Matching Pursuit \citep{Mallat1993}, which is conceptually easier to implement, is also a suitable and efficient algorithm to solve the task.
 
Recovery of the sparse (or compressible) signal $\mathbf{x}$ is guaranteed providing that the correlation between any two elements of the dictionary is small (as measured by the mutual coherence indicator or the K-restricted isometry constant) and the number of measurements is large enough \citep{Candes2006}. In case of overcomplete and redundant dictionaries, the successful restoration of the signal relies on the use of a prior as it is often the case for solving many inverse problems, with the maximum a posteriori (MAP) estimator for instance. The prior we are interested in is the sparsity of the solution we are looking for, and it can be introduced following two approaches which are closely related but not equivalent for such redundant dictionaries, as studied by \citet{Elad2007}.

The first approach relies on an analysis-based prior. The signal $\mathbf{x}$ characterized by its inner products with all the atoms of a dictionary $\mathbf{A}$ is assumed to be sparse for this dictionary, that is $\boldsymbol{\gamma}_{\rm{a}} = \mathbf{A}\!^\top\mathbf{x}$ with $\boldsymbol{\gamma}_{\rm{a}}$ the sparse representation of $\mathbf{x}$. To be efficient, this approach must involve priors on the signal for selecting adequate dictionaries like for instance wavelet-based ones for nearly-isotropic sources in astronomical images, as it was implemented in the previous algorithm \texttt{OV\_WAV}. Given $\ell_{1}(\mathbf{x})=\sum_{i} \left|\mathbf{x}_{i}\right|$, equation (\ref{eq:general_solution_inverse_problem}) becomes:

\begin{equation}
    \label{eq:analysis_solution}
    \hat{\mathbf{x}} = \argmin_{\mathbf{x'}} \frac{1}{2} \norm{\mathbf{
y} - \mathbf{\mathcal{H} x'}}^{2}_{2} + \lambda \ell_{1}(\mathbf{A}\!^{\top}\mathbf{x'}).
\end{equation}

\noindent The second approach is sparse synthesis, where the signal to be restored $\mathbf{x}$ is assumed to be a linear combination of a few atoms from a dictionary, so that $\mathbf{x}= \mathbf{S}\boldsymbol{\gamma}_{\rm{}s}$ where $\gamma_{\rm{}s}$ is the sparse representation of $\mathbf{x}$ and $\mathbf{S}$ the synthesis dictionary (not to confound with the measure of sparsity $\mathcal{S}$). That leads to write the solution of the inverse problem as:

\begin{equation}
    \label{eq:synthesis_solution}
    \hat{\mathbf{x}} = \argmin_{\mathbf{x'}} \frac{1}{2} \norm{\mathbf{
y} - \mathbf{\mathcal{H} \mathbf{S}}\boldsymbol{\gamma}_{\rm{s}}}^{2}_{2} + \lambda \ell_{1}(\boldsymbol{{\gamma}}_{\rm{s}}).
\end{equation}

For redundant dictionaries, solutions for analysis or synthesis priors are different. As far as the authors know, no general results on their practical comparison are available for usual transforms, even for the $\ell_{1}$ norm case. However, the analysis approach may be more robust than the synthesis approach as it does not require for the signal to be expressed as a linear combination of atoms of a given dictionary. It is clear that a synthesis approach with dictionaries including too few atoms leads to a rough restoration and that the number of unknowns for large dictionaries is computationally expansive and often prohibitive.

\subsection{\texttt{DAWIS}: A synthesis-by-analysis approach}
\label{subsec:synthesis_by_analysis}

We chose for \texttt{DAWIS} a hybrid approach, the analysis-by-synthesis method first developed for processing radio astronomy images and explicitly implemented in the \texttt{MORESANE} algorithm \citep{Dabbech2015}. The principle is to model an image as a linear combination of synthesis atoms that are learned iteratively through analysis-based priors. The previous algorithm, \texttt{OV\_WAV}, already follows this path implicitly since i)~it makes use of wavelet dictionaries to reconstruct images according to an analysis approach where objects are detected and restored subject to the wavelet coefficient values, and ii)~it sometimes has to be run iteratively two or three times to obtain better results, hence the beginning of a synthesis approach
with the successive restored images as synthesis atoms. However, applying a formal analysis-by-synthesis method allows to solve most of the problems met by the \texttt{OV\_WAV} algorithm while keeping the advantages of an analysis based on wavelet atoms for the detection of low surface brightness features.

This approach makes \texttt{DAWIS} (like \texttt{MORESANE}) a very versatile tool with a great range of applications, as the nature of the signal of interest and of the degradation operator are defined by the person performing the analysis. In the work presented here for example, the signal to be recovered is the ICL and the degradation operator $\mathbf{\mathcal{H}}$ can then be seen as effects coming from both instrumental (scattered light, PSF) and physical (blending of astronomical sources, contamination by diffuse halos, etc.) origins. However, our goal utilizing \texttt{DAWIS} here is not to cope with instrumental effects (this kind of degradation must therefore have been dealt with before) but to focus on both the detection of diffuse and low surface brightness features (e.g., detecting signal where standard detection methods fail) and source disentangling (e.g., separating the ICL from the galaxies based on non-arbitrary parameters). The operating mode of \texttt{DAWIS} is therefore conditioned by these specific analysis goals. We stress the fact that other applications are possible (also see Section~\ref{subsec:atom_classification} for more details).

\subsection{\texttt{DAWIS}: A semi-greedy algorithm}
\label{subsec:semi_greedy}

The synthesis-by-analysis approach implemented in \texttt{MORESANE} by \citet{Dabbech2015} and in \texttt{DAWIS} by us is conceptually reminiscent of a Matching Pursuit algorithm \citep{Mallat1993} where the atoms of the dictionary are constructed with orthogonal projections of the signal on time-frequency functions. At each iteration the best correlated projection is kept as an atom for the synthesis dictionary, and retrieved from the signal before applying the same process on the residual. Such a method is efficient in determining atoms that characterize well the signal since its core strategy is to minimize at each iteration a residual from the computation of the inner products of the under-construction solution with all the atoms of the analysis dictionary, hence a so-called greedy method which has the main disadvantage to be time consuming. A well-known greedy algorithm is \texttt{CLEAN}, which uses a set of Diracs and a PSF to deconvolve the image of a field of stars represented by intensity peaks. The algorithm assesses to each peak a Dirac function with a spatial position in the image and an amplitude, which makes it sparse by nature as the observed field is expressed with few positive coefficients. The sky image is recovered with the convolution of the set of Diracs by the PSF, which can be seen as the product of coordinates by a dictionary with one single atom. 

A problem when detecting sources in astronomical images is blending: a faint source aside a much brighter one is partially hidden by the bright source, especially if the latter is also larger. In \texttt{CLEAN}, an iterative process controlled by an empirical factor called the \texttt{CLEAN} factor is introduced to address this issue. The highest detected peak in the image is first considered and, instead of totally removing it from the image before computing and processing the new highest peak, only a fraction of it given by this \texttt{CLEAN} factor is subtracted. By doing so, the risk of accidentally removing faint blended sources is decreased, hence allowing a better deconvolution of the image.

In \texttt{DAWIS} the brightest source within the image is detected at each iteration in the wavelet space before being restored and removed from the image. Likewise, a very bright object in the image (such as a foreground star) translates into wavelet coefficients with very high values which strongly dominate this representation space (even at low frequency scales), and contaminates low surface brightness features. Removing this bright structure first before detecting fainter sources allows a better recovery of the objects compared to the previous process used in the \texttt{OV\_WAV} algorithm. Similarly to \texttt{MORESANE}, we also introduced in \texttt{DAWIS} a parameter $\delta\in[0,1]$ equivalent to the \texttt{CLEAN} factor so that only a fraction of the reconstructed object is actually removed from the image, limiting also the appearance of artifacts.

The downside of such an approach is its slowness, since an analysis where sources are processed one by one is way too computationally time expensive. This is why we decided for \texttt{DAWIS} to implement \texttt{MORESANE}'s semi-greedy method. The reconstructed atom at each iteration is not composed of a single source, but of a set of sources with similar characteristic sizes and intensities. This set is defined by a parameter $\tau\in[0,1]$, setting a threshold relative to the brightest structure in the image.

\subsection{\texttt{DAWIS}: the B3-spline wavelet as analysis dictionary}
\label{subsec:IUWT}

As previously said, an analysis-based method is used to obtain the synthesized atoms of $\mathbf{S}$. As ICL is believed to have an isotropic or quasi-isotropic shape, we chose as adequate dictionary the well known B3-spline wavelet one, with $\mathbf{w}=\mathbf{A}\!^{\top}\mathbf{x}$ where $\mathbf{w}$ is the vector of wavelet coefficients. The choice of such a symmetric and compact mother wavelet grants also a very efficient way to compute these coefficients using the IUWT, for which there is no need to compute the product between $\mathbf{A}\!^{\top}$ and $\mathbf{x}$.
Taking benefit of the so-called \textit{\`a trous} algorithm \citep{Holschneider1989}, the original image $\mathbf{x}$ is smoothed consecutively $J$ times using an adaptive B3-spline kernel giving $J$ coarse versions of $\mathbf{x}$ with $\mathbf{c}^{(j)}\in\mathbb{R}^{N}$ being the version at the scale $j$. The vector $\mathbf{w}$ can then be written as the concatenation of $J+1$ vectors $\mathbf{w}^{(j)}\in\mathbb{R}^{N}$ such as $\mathbf{w}=\{\mathbf{w}^{(1)}|...|\mathbf{w}^{(j)}|...|\mathbf{w}^{(J)}|\mathbf{c}^{(J)}\}$, where $\mathbf{w}^{(j)}=\mathbf{c}^{(j+1)}-\mathbf{c}^{(j)}$ \citep{Mallat1989, Shensa1992}. Each vector $\mathbf{w}^{(j)}$ basically represents the details between two consecutive smoothed levels, and the components of those vectors are called the wavelet coefficients. An image of size $N$ pixels gives a maximal number of scales $J\leq\log_{2}(N)-1$.

\subsection{\texttt{DAWIS}: Noise filtering and multiresolution support}
\label{subsec:filtering_support}

Determining which wavelet coefficients are of interest (filtering step) requires to know how the noise in the direct space translates in the wavelet space. Since the wavelet transform is a linear one, the noise statistics remain the same. To be able to select wavelet coefficients using a simple thresholding method involving Gaussian statistics, one needs to ensure such a Gaussian distribution for the noise in the wavelet space. For this purpose \texttt{DAWIS} makes use of a variance stabilization transform. Considering an image $\mathbf{x}$ with a combination of Gaussian noise and Poissonian noise, \texttt{DAWIS} involves the generalized Anscombe transform $\mathcal{A}:\mathbb{R}^{N}\to\mathbb{R}^{N}$ \citep{Anscombe1948}, which is given by:

\begin{equation}
    \label{eq:anscombe_transform}
    \mathcal{A}(\mathbf{x})=\frac{2}{g}\left( g \mathbf{x} + \frac{3}{8} g^{2} + \sigma^{2} - g \mu \right)^{\frac{1}{2}},
\end{equation}

\noindent
with $g$ being the gain of the detector, and $\mu$ and $\sigma$ being respectively the mean and standard deviation of the Poissonian-Gaussian noise in the original image, computed here with a bisection-like method. The result $\Tilde{\mathbf{x}}=\mathcal{A}(\mathbf{x})$ is an image with a Gaussian noise of $\sigma=1$, which has a very nice behaviour in the wavelet space. The \textit{\`a trous} algorithm can then be applied to the output image to create the stabilized wavelet coefficient vectors $\Tilde{\mathbf{w}}^{(j)}$.

The statistically significant pixels are then selected at each scale using a thresholding method and a multiscale support is identified \citep{Bijaoui1995}, which is given a scalar operator $\mathcal{T}$, a vector $\mathbf{M}=\{\mathbf{M}^{(1)}|...|\mathbf{M}^{(j)}|...|\mathbf{M}^{(J)}\}$ with $\mathbf{M}^{(j)}\in\mathbb{R}^{N}$ such as:

\begin{equation}
    \label{eq:hamultiscale_support}
    \mathbf{M}_{i}^{(j)}= \left\{
    \begin{array}{ll}
        1 & \mbox{if } \mathcal{T}(\Tilde{\mathbf{w}}_{i}^{(j)}) > 0 \\
        0 & \mbox{otherwise.}
    \end{array}
\right.
\end{equation}

\noindent
The computation of the multiscale support fills two objectives. First it acts as a $\ell_{0}$ sparsification of the analysis coefficients, discarding small (non-significant) wavelet coefficients and indicating the position of interesting features. Secondly, it allows us to translate in a very easy way this acquired knowledge from the variance stabilized wavelet space to the non-stabilized one (e.g. the vector $\mathbf{w}$ generated by $\mathbf{A}\!^{\top}\mathbf{x}$), where the actual object identification is done.

Since the noise in the initial image is considered to be spatially uncorrelated (which is not always the case in reality) and therefore generates wavelet coefficients with high values only at the first two high frequency scales, a first relevant approach is to estimate the standard deviation of the noise and its mean at those two first scales. The IUWT being a linear transform, the rms values can then be extrapolated to the higher wavelet scales where i) the noise becomes highly-correlated due to the large size of the related filters and ii) the mean size of sources increases, leaving less and less background pixels to estimate the noise statistics. 

The threshold operator applied to the wavelet coefficients $\mathcal{T}$ can take many forms. \texttt{DAWIS} implements the usual hard threshold operator which operates as:

\begin{equation}
    \label{eq:hard_threshold}
    \mathcal{T}(\Tilde{\mathbf{w}}_{i}^{(j)}) = \left\{
    \begin{array}{ll}
        \Tilde{\mathbf{w}}_{i}^{(j)} & \mbox{if } |\Tilde{\mathbf{w}}_{i}^{(j)}| \geq t \\
        0 & \mbox{otherwise.}
    \end{array}
    \right.
\end{equation}

\noindent
The threshold $t$ applied to each $\Tilde{\mathbf{w}}^{(j)}$ is different, such as $t=k\sigma^{(j)}$, with $\sigma^{(j)}$ being the standard deviation of the noise of $\Tilde{\mathbf{w}}^{(j)}$ and $k$ a constant usually chosen to be 3 or 5 according to the chosen probability for false alarms. Other formulations for $\mathcal{T}$ can be used, such as the soft threshold operator \citep{Mallat2008} or the combined evidence operator used in \texttt{OV\_WAV} \citep{DaRocha2005}.

\subsection{\texttt{DAWIS}: Object identification via interscale connectivity}
\label{subsec:object_identification}

Let us first repeat that a source in the original image generates significant wavelet coefficients at several successive scales for a non orthogonal transform. Consequently, an analysis along the scale axis has to be performed in order to identify the set of wavelet coefficients related to this source. To this end, \texttt{DAWIS} follows once again the recipe first proposed by \citet{Bijaoui1995} and also implemented in \texttt{MORESANE}. It relies on the construction of interscale  trees.

Let us denote $\boldsymbol{\alpha}$ the set of significant wavelet coefficients of $\mathbf{w}$. The location of these coefficients is given by the multiscale support $\mathbf{M}$ defined in Section~\ref{subsec:filtering_support}. When applying this mask, a classical segmentation procedure is first performed to group these coefficients into regions (e.g. domains) of connected pixels. One can then write $\boldsymbol{\alpha}$ as a concatenation of vectors $\boldsymbol{\alpha}^{(j)}$ so that $\boldsymbol{\alpha}=\{\boldsymbol{\alpha}^{(1)}|...|\boldsymbol{\alpha}^{(j)}|...|\boldsymbol{\alpha}^{(J)}\}$ with $\boldsymbol{\alpha}^{(j)}\subset\mathbf{w}^{(j)}$, each $\boldsymbol{\alpha}^{(j)}$ being composed of a set of domains $\mathbf{d}$ with different sizes.
As usual, each significant region is characterized by the location and amplitude of its  coefficient with the highest value, allowing one to define a local maximum. Let $\mathbf{w}^{(j)}_{i}$ be this local maximum value for the region $\mathbf{d}_{(1)}\subset\boldsymbol{\alpha}^{(j)}$. Region $\mathbf{d}_{(1)}$ is then defined as linked to a region $\mathbf{d}_{(2)}\subset\boldsymbol{\alpha}^{(j+1)}$ if $\mathbf{w}^{(j+1)}_{i}$ belongs to $\mathbf{d}_{(2)}$. Finally, by testing such a connectivity for each region $\mathbf{d}\subset\boldsymbol{\alpha}$, interscale trees are built through the whole segmented wavelet space.

With this procedure, an object $\mathbf{O}$ is then defined by the concatenation of $K$ connected regions in the wavelet space $\mathbf{d}_{(k)}$ such as $\mathbf{O}=\{\mathbf{d}_{(1)}|...|\mathbf{d}_{(k)}|...|\mathbf{d}_{(K)}\}$. Note that a region is linked to at most one other region at the next lower frequency scale, but can be linked to several regions at the next higher frequency scale. One criterion must be satisfied to consider that such a tree is related to a genuine object in the direct image, and not an artifact. An interscale tree must include at least three regions linked together at three successive scales.

\subsection{\texttt{DAWIS}: Object reconstruction}
\label{subsec:object_reconstruction}
Applying the procedure summarized in the previous section, each object in the direct space is therefore related to an interscale tree of significant wavelet coefficients. Beyond the structure of the tree itself, the amount of information about this object is distributed across the linked regions and can be measured by the value of their maxima once normalized. Since wavelets act as a contrast detector, the average values of the wavelet coefficient amplitudes tend to be indeed lower at small scales than at large scales in astronomical images, thereby creating an implicit bias when comparing wavelet coefficient values at different scales. As a normalization factor for each scale $j$, \texttt{DAWIS} uses the standard deviation at this scale $j$ of the IUWT of a Gaussian white noise image with unit variance, denoted $\sigma_{1}^{(j)}$. For an object $\mathbf{O}$, this results in the normalized vector $\Tilde{\boldsymbol{\alpha}}$ of size $K$,  with given $\mathbf{d}_{(k)}\subset\boldsymbol{\alpha}^{(j)}$, $\Tilde{\boldsymbol{\alpha}}_{k}=\max\frac{\mathbf{d}_{(k)}}{\sigma_{1}^{(j)}}$.

Especially relevant for the information content is the region $\mathbf{d}_{(k)}\subset\mathbf{O}$ in the tree which contains most of the information about the object. This region of maximum information is the interscale maximum. It is denoted by an index $k_{\rm{obj}}$ and a scale $j_{\rm{obj}}$ such as:

\begin{equation}
    k_{\rm{obj}}=\argmax_{k} \Tilde{\boldsymbol{\alpha}}_{k}.
\end{equation}

\noindent
This interscale maximum gives an easy way to characterize an object $\mathbf{O}$ since the parameter $j_{\rm{obj}}$ provides its characteristic size $2^{j_{\rm{obj}}}$. Moreover, the parameter $\Tilde{\boldsymbol{\alpha}}_{k_{\rm{obj}}}$ giving its normalized intensity is also used by \texttt{DAWIS} to compare it to other objects and rank them for the restoration step. As indicated in Section~\ref{subsec:semi_greedy}, a parameter $\tau$ defines a threshold relative to the brightest identified object, the one with the highest parameter $\Tilde{\boldsymbol{\alpha}}_{k_{\rm{obj}}}$, denoted $\Tilde{\boldsymbol{\alpha}}_{\rm{max}}$. Only objects verifying $\Tilde{\boldsymbol{\alpha}}_{k_{\rm{obj}}}\geq \tau \Tilde{\boldsymbol{\alpha}}_{\rm{max}}$ are restored at a given iteration of the processing and are included in the associated dictionary atom.

Objects are reconstructed individually, using their specific support $\mathbf{M}_{\rm{spec}}$ giving the location in $\mathbf{w}$ of every significant coefficient belonging to $\mathbf{O}$. Note that here \texttt{DAWIS} also strictly follows the procedure from \citep{Bijaoui1995} and that regions at scales $j$ higher than $j_{\rm{obj}}$ are discarded for the restoration. This restoration is a direct application of Equation~(\ref{eq:general_solution_inverse_problem}), where positivity of the solution is used as a regularization term, with $\mathbf{y}=\mathbf{M}_{\rm{spec}}\mathbf{w}$ and $\mathcal{H}=\mathbf{M}_{\rm{spec}}\mathbf{A}\!^{\top}$. \texttt{DAWIS} uses the conjugate gradient algorithm from \citet{Bijaoui1995}, which makes use of the adjoint operator $\mathbf{^{\dag}{A}}$ of the analysis dictionary $\mathbf{A}$ (see their article for an explicit definition of this operator). This algorithm iteratively finds the solution $\hat{\mathbf{x}}$, which is the reconstructed object. Note that this conjugate gradient version algorithm makes use of the Fletcher-Reeves step size $\beta$ \citep{Fletcher1964}. This process is applied to all objects, before concatenating them into a single restored image $\mathbf{z}\in\mathbb{R}^{N}$ such as $\mathbf{z}=\sum_{l}\hat{\mathbf{x}}_{(l)}$.

\subsection{\texttt{DAWIS}: Architecture of the algorithm}
\label{subsec:architecture}

We give in this section the general architecture of \texttt{DAWIS} in the form of a simplified pseudo-code: Algorithms~\ref{algo:DAWIS_main} and \ref{algo:DAWIS_atom_synthesis} summarize the synthesis-by-analysis approach as explained in Sections~\ref{subsec:synthesis_by_analysis} and \ref{subsec:semi_greedy}, while Algorithms~\ref{algo:DAWIS_multiscale_support}, \ref{algo:DAWIS_interscale_analysis} and \ref{algo:conjugate_gradient_algorithm} describe the wavelet atom based analysis as described in Sections~\ref{subsec:IUWT} to \ref{subsec:object_reconstruction}. Concerning the operating mode of \texttt{DAWIS} which is iterative, some parameters have to be defined to control the convergence of the algorithm.

To ensure that the algorithm is correctly peeling the image starting by the bright sources, we impose an upper scale $J$ so that the brightest detected object at each iteration cannot have an interscale maximum $\Tilde{\boldsymbol{\alpha}}_{\rm{max}}$ at scale $j_{\rm{max}}>J$. This also decreases the computation time, as for the first few iterations there is no need to perform the wavelet and interscale analyses for all scales. The upper scale is initialized at $J=3$ since an interscale tree needs at least connected domains on three successive scales to be considered as related to an object.

The main convergence parameter is defined as $\frac{\norm{\sigma_{(i-1)} - \sigma_{(i)}}}{\sigma_{(i-1)}}$ and is computed from the variation of the standard deviation of the residual at each iteration $i$. However the nature of the synthesis atom for different iterations can induce instability for this convergence parameter. One atom can indeed be composed of many bright objects, which means strong variation of the standard deviation, and  another of a single faint object, hence a low variation of the standard deviation which might break the loop while there are still sources in the residual. Therefore we normalize this convergence parameter by  the number of objects $n_{\rm{obj}}$ to stabilize it. Once the value of this parameter goes below a threshold $\epsilon$, the value of the upper scale $J$ is increased by one, so that larger sources can be processed by the algorithm. A hard limit $N_{\rm{it}}$ is also given to the algorithm to restrict the number of possible iterations.

\normalem 
\begin{algorithm}
\caption{\texttt{DAWIS} - main algorithm}
\label{algo:DAWIS_main} 

\DontPrintSemicolon
  
\KwInput{$\tau$, $\delta$, $\epsilon$, $N_{\rm{it}}$, original image $\mathbf{x}$ with size $N$.}
\KwOutput{synthesised dictionary $\mathbf{S}$, restored image $\mathbf{z}^{\rm{tot}}$, residual $\mathbf{r}$.}

Initialize $J = 3, i = 1, \mathbf{r}_{(0)}=\mathbf{x}$.\;

\While{$J < \log_{2}(N) - 1$}
{
   	\While{$i \le N_{\rm{it}} $}
    {
        $i=i+1$.\;
        Compute $\mathbf{w}$ and $\mathbf{M}$ with \textbf{Algorithm~\ref{algo:DAWIS_multiscale_support}}.\;
        Compute $\{\mathbf{O}_{(1)}|...|\mathbf{O}_{(n)}\}$ and $\{\Tilde{\alpha}_{(1)}|...|\Tilde{\alpha}_{(n)}\}$ with \textbf{Algorithm~\ref{algo:DAWIS_interscale_analysis}}.\;
   		Compute $\mathbf{z}_{(i)}$ and $n_{\rm{obj}}$ with \textbf{Algorithm~\ref{algo:DAWIS_atom_synthesis}}.\;
   		
   		Update dictionary $\mathbf{S} = \{ \mathbf{S}~|~\mathbf{z}_{(i)}\}$.\;
   		Update residuals $\mathbf{r}_{(i)} = \mathbf{r}_{(i-1)} - \mathbf{z}_{(i)}$.\;
   		Update restored image $\mathbf{z}^{\rm{tot}} = \mathbf{z}^{\rm{tot}} + \mathbf{z}_{(i)}$.\;
   		Compute $\sigma_{(i)}$ standard deviation of $\mathbf{r}_{(i)}$.\;
   		\If{$\frac{\norm{\sigma_{(i-1)} - \sigma_{(i)}}}{ n_{\rm{obj}}\sigma_{(i-1)}} \leq \epsilon$}
        {
            Break loop on $i$.\;
        }
    }
    $J = J + 1$\;
}

Final residual $\mathbf{r} = \mathbf{r}_{(i)}$.\;
\end{algorithm}


\begin{algorithm}
\caption{\texttt{DAWIS} - wavelet analysis \& multiscale support}
\label{algo:DAWIS_multiscale_support} 

\DontPrintSemicolon
 
\KwInput{Maximum scale $J$, residual $\mathbf{r}$.}
\KwOutput{wavelet coefficients $\mathbf{w}$ and multiscale support $\mathbf{M}$.}

Apply Anscombe transform $\Tilde{\mathbf{r}}=\mathcal{A}(\mathbf{r})$.\;
Compute $\mathbf{\Tilde{w}}=\mathbf{A}\!^{\top}\Tilde{\mathbf{r}}$ with \textit{\`a trous} algorithm for scale $j=1$ to $J$.\;
Apply hard threshold $\mathbf{\Tilde{w}} = \mathcal{T}(\mathbf{\Tilde{w}})$.\;
Compute multiscale support $\mathbf{M}$ from $\mathbf{\Tilde{w}}$.\;
Compute $\mathbf{w}=\mathbf{A}\!^{\top}\mathbf{r}$ with \textit{\`a trous} algorithm for scale $j=1$ to $J$.\;

\end{algorithm}

\begin{algorithm}
\caption{\texttt{DAWIS} - interscale analysis}
\label{algo:DAWIS_interscale_analysis} 

\DontPrintSemicolon
 
\KwInput{wavelet coefficients $\mathbf{w}$, multiscale support $\mathbf{M}$.}
\KwOutput{objects $\{\mathbf{O}_{(1)}|...|\mathbf{O}_{(n)}\}$, interscale maximums $\{\Tilde{\alpha}_{(1)}|...|\Tilde{\alpha}_{(n)}\}$.}

Initialize $m=1$, $n=1$.\;
Compute domains $\mathbf{d}=\{\mathbf{d}_{(1)}|...|\mathbf{d}_{(M)}\}$ from $\mathbf{w}$ and $\mathbf{M}$ with \textit{segmentation} algorithm.\;

\While{$m \leq M$}
{
    Compute spatial position $i_{\rm{max}}$ and scale $j_{\rm{max}}$ of local maximum $\max(\mathbf{d}_{(m)})$.\;
    \If{$\mathbf{w}^{(j_{\rm{max}}+1)}_{i_{\rm{max}}}\in\mathbf{d}_{m'}$}
    {
        Link $\mathbf{d}_{m}$ and its linked domains to $\mathbf{d}_{m'}$.\;
    }
    \Else
    {
        Define interscale tree $\mathbf{O}_{(n)}$ as $\mathbf{d}_{(m)}$ and its linked domains.\;
        \If{$\mathbf{O}_{(n)}~\rm{has~at~least~3~linked~domains}$}
        {
            Compute interscale maximum $\Tilde{\alpha}_{(n)}$ as in Section~\ref{subsec:object_reconstruction}.\;
            Update objects $ \{\mathbf{O}_{(1)}~|...|~\mathbf{O}_{(n)}\}$.\;
            Update interscale maximums $\{\Tilde{\alpha}_{(1)}~|...|~\Tilde{\alpha}_{(n)}\}$.\;
            $n=n+1$.\;
        }
    }
    $m=m+1$.\;
    
}

\end{algorithm}
\begin{algorithm}
\caption{\texttt{DAWIS} - atom synthesis}
\label{algo:DAWIS_atom_synthesis} 

\DontPrintSemicolon
 
\KwInput{$\tau$, $\delta$, objects $\{\mathbf{O}_{(1)}|...|\mathbf{O}_{(n)}\}$, interscale maximums $\{\Tilde{\alpha}_{(1)}|...|\Tilde{\alpha}_{(n)}\}$.}
\KwOutput{synthesised atom $\mathbf{z}$ and $n_{\rm{obj}}$ number of objects in it.}


Initialize $n_{\rm{obj}}=0$, $l=1$.\;
Brightest object $\Tilde{\alpha}_{\rm{max}} = \max(\{\Tilde{\alpha}_{(1)}|...|\Tilde{\alpha}_{(n)}\})$.\;

\While{$l \leq n$}
{
    \If{$\Tilde{\alpha}_{(l)}\geq \tau \Tilde{\alpha}_{\rm{max}}$}
    {
    $n_{\rm{obj}}=n_{\rm{obj}}+1$.\;
    Compute restored object $\hat{\mathbf{x}}_{(l)}$ with Algorithm~ \ref{algo:conjugate_gradient_algorithm}.\;
    Update $\mathbf{z}=\mathbf{z} + \hat{\mathbf{x}}_{(l)}$.\;
    
    }
    $l = l + 1$.\;
}
Apply \texttt{CLEAN} factor $\mathbf{z}=\delta \mathbf{z} $.

\end{algorithm}


\begin{algorithm}
\caption{Conjugate gradient algorithm}
\label{algo:conjugate_gradient_algorithm} 

\DontPrintSemicolon
  
\KwInput{Flux convergence parameter $\epsilon$, original object image $\mathbf{x}$, object wavelet coefficients $\mathbf{y}=\mathbf{M}_{\mathrm{spec}}\mathbf{w}$, maximum number of iterations $N_{\mathrm{iter}}$.}

\KwOutput{Restored object image $\mathbf{\Tilde{x}}$.}

Initialize $k = 0$.\\
Initialize $\mathbf{\Tilde{x}}^{(0)} = \mathbf{^{\dag}\mathbf{A}}\mathbf{y}$.\\
Initialize $\mathbf{w}_{r}^{(0)} = \mathbf{y} - \mathbf{A} \mathbf{\Tilde{x}}^{(0)}$.\\
Initialize $\mathbf{r}^{(0)} = \mathbf{v}^{(0)} =~^{\dag}\mathbf{A} \mathbf{w}_{r}^{(0)}$.\\

\While{$k<N_{\mathrm{iter}}$}
{
Compute the step size $\delta^{(k)} = \norm{\mathbf{r}^{(k)}} / \norm{\mathbf{w}_{r}^{(k)}}$.

Update the restored image $\mathbf{\Tilde{x}}^{(k+1)} = \mathbf{\Tilde{x}}^{(k)} + \delta^{(k)} \mathbf{v}^{(k)}$.

Set negative coefficients of $\mathbf{\Tilde{x}}^{(k+1)}$ to 0 (positivity of the solution).\\

Update the wavelet residuals $\mathbf{w}_{r}^{(k+1)} = \mathbf{y} - \mathbf{A} \mathbf{v}^{(k)}$.\\

Update the direct space residual $\mathbf{r}^{(k+1)} =~^{\dag} \mathbf{A} \mathbf{w}_{r}^{(k+1)}$.\\

Compute the step size $\beta^{(k)} = \frac{\norm{\mathbf{r}^{(k+1)}}^2_2}{\norm{\mathbf{r}^{(k)}}^2_2}$.\\

Compute the new conjugate direction $\mathbf{v}^{(k+1)} = \mathbf{r}^{(k+1)} + \beta^{(k)} \mathbf{r}^{(k)}$.\\

\If{$\frac{\norm{\mathbf{\Tilde{x}}^{(k+1)} - \mathbf{\Tilde{x}}^{(k)}}}{\norm{\mathbf{\Tilde{x}}^{(k)}}} \le \epsilon$}
{
Set final restored object image $\mathbf{\Tilde{x}}=\mathbf{\Tilde{x}}^{(k+1)}$.\\
End loop on $k$.\\
}
Set $k = k + 1$.\\
}
\end{algorithm}

\ULforem 

\subsection{\texttt{DAWIS}: implementation and parallelization}
\label{dawis_implementation}

\texttt{DAWIS} is implemented with an emphasis on modularity (e.g., a set of modules that can be moved or replaced by new versions), as the algorithm can still be upgraded to increase the quality of the method or modified according to new analysis goals. We chose to write the main layer of modules in Python, as it is a very commonly used, versatile and accessible open-source language. However this versatility has a cost in terms of numerical performance, which led us to support the Python modules by Fortran 90 codes where the main numerical computations are done.

As explained in Section~\ref{subsec:implementation_and_limitations}, one of the main limitations of wavelet based algorithms is computation time, which prevents the application of previous packages like \texttt{OV\_WAV} on large samples of images or to very large images. A big effort has been made with \texttt{DAWIS} to parallelize the algorithm. This is not straightforward, as the main algorithm is iterative, which means that only the content of one iteration can be speed-up. Therefore, we parallelized the modules inside an iteration. When working on large data arrays (which is typically the case both when computing the wavelet data cube and the multiresolution support with Algorithm~\ref{algo:DAWIS_multiscale_support} and when performing the multiscale analysis with Algorithm~\ref{algo:DAWIS_interscale_analysis}), the Fortran modules are parallelized in shared memory using \texttt{OpenMP}\footnote{https://www.openmp.org/}. Complementary, we use the new Python package \texttt{Ray}\footnote{https://ray.readthedocs.io/en/latest/} to distribute processes when working on many small arrays (the restorations of numerous objects to compute the associated synthesis atom are independent from each other and can be distributed, such as Algorithm~\ref{algo:DAWIS_atom_synthesis} and Algorithm~\ref{algo:conjugate_gradient_algorithm} for example).

\begin{figure}
    \centering
    \includegraphics[width=\hsize]{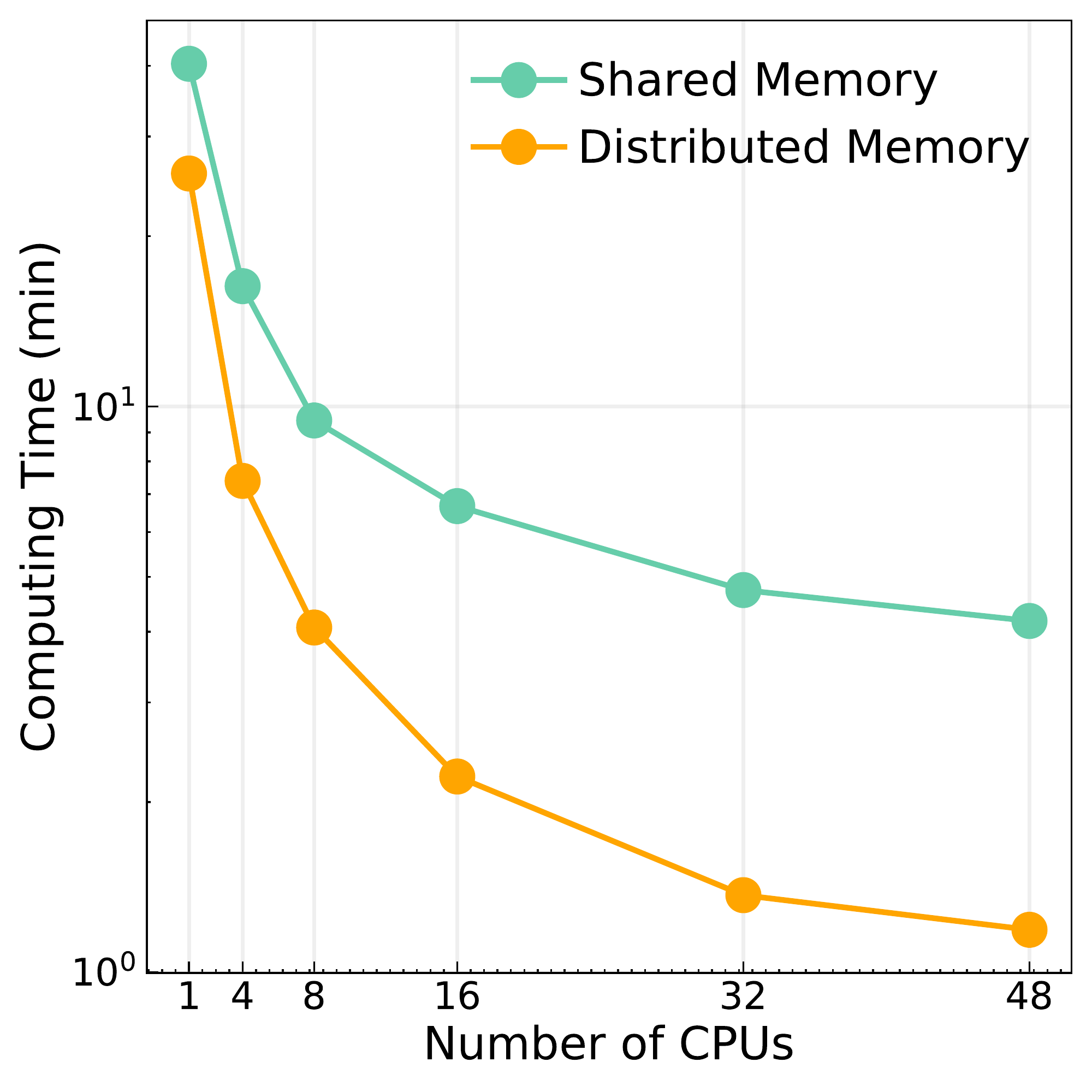}
    \caption{CPU computing time scaling test for the two kinds of parallelization used in \texttt{DAWIS}. The scale for the computing time is logarithmic. Notice that this gain of time is achieved for the modules inside one main iteration of the algorithm, and does not represents the total CPU computing time of \texttt{DAWIS} (see the text for more details).}
    \label{fig:plot_para_test}
\end{figure}

We display here a CPU computing time scaling test on mock data for both types of parallelizations. For the shared memory parallelization, the test is set on an image of size $4096\times4096$ pixels (giving a wavelet data cube of $4096\times4096\times10$ wavelet coefficients). Algorithm~\ref{algo:DAWIS_multiscale_support} and Algorithm~\ref{algo:DAWIS_interscale_analysis} are run on the image first serially (one CPU), and then by increasing progressively the number of CPU. As shown in Figure~\ref{fig:plot_para_test}, the gain of computing time is high when increasing the number of CPUs from 1 to 16, going from a computing time of $\sim$~40 minutes to a computing time of $\sim$~6 minutes. However, these modules do not scale linearly with the number of CPUs, and the gain in computing time is rather negligible for 32 and more CPUs, where the computing time is converging toward a value of $\sim$~4 minutes.

For the distributed memory parallelization, the test is run on 1000 copies of the same object array of size $128\times128$ pixels. Algorithm~\ref{algo:DAWIS_atom_synthesis} and Algorithm~\ref{algo:conjugate_gradient_algorithm} are run on the arrays, and  the resulting CPU computing times are also displayed in Figure~\ref{fig:plot_para_test}. Similar to the shared memory parallelization, the gain in CPU computing time is mostly impactful when increasing to 16 the number of CPUs with a computing time value going from $\sim$~25 minutes to $\sim$~2 minutes. It is however rather negligible with higher numbers of CPUs, and converges toward a computing time value of $\sim$~2 minutes. Of course, the objects to restore are not always of size $128\times128$ pixels, and the number of objects to restore also can differ from 1000. In some cases, the sizes of objects can even potentially be a large fraction of the image. In such cases however, there are rarely more than one or two objects to restore. In our experience, the computation time of an iteration with one or two large objects to restore does not differ very much from the computation time of an iteration where many small objects are restored.

Notice that this simple test is not representative of the performance of the complete algorithm. Indeed, due to the greedy nature of Algorithm~\ref{algo:DAWIS_main}, the CPU computing time of \texttt{DAWIS} depends largely on the content of the image. Indeed, an image containing complex structures will always be longer to process than an image with very simple shapes, as more main iterations are needed to accurately model these complex structures (in our experience, the number of main iterations ranges from a hundred to few hundreds, depending on the image). Additionally, the length of a main iteration also greatly depends on the content of the image. For example, the algorithm will sometimes choose to restore only one very bright source. In that case, the parallelization in distributed memory is of no use, as there is only one object to restore. Nevertheless, these parallelization processes ensure that the complete CPU computing time of \texttt{DAWIS} does not disproportionately increase in case of 'heavy' main iterations.

\subsection{Astrophysical priors on object selection}
\label{subsec:atom_classification}

As written in Section~\ref{subsec:synthesis_by_analysis}, the columns of the final synthesized dictionary $\mathbf{S}$ are atoms which are no more than the restored images $\mathbf{z}_{i}$ at each iteration $i$. Then, the restoration $\mathbf{z}^{\rm{tot}}$ of the whole original image is given by the sum of all these atoms such as $\mathbf{z}^{\rm{tot}}=\sum_{i}\mathbf{z}_{(i)}$. The generic operating mode of \texttt{DAWIS} therefore produces a fully (denoised) restored image every time the algorithm is run. However, in this way a large part of the information recovered by the sparse synthesis-by-analysis method is not utilized, since all synthesized atoms are concatenated into the same image in which the information on individual spatial frequencies or characteristic sizes is no longer accessible. However, depending on the analysis goals, one might be only interested in subsets of atoms of $\mathbf{S}$ or, even more specifically, one might want to compose alternative synthesized dictionaries, the atoms of which would be selected differently throughout the synthesis-by-analysis procedure. To allow for such possibilities, a discrimination operator $\mathcal{D}$ is applied to the detected objects before constructing the associated synthesis atoms such as $\mathbf{z}'=\sum_{l}\mathcal{D}\,\,\hat{\mathbf{x}}_{(l)}$. We will not give here a rigorous definition of such an operator, as it can take many forms and use different properties to discriminate objects depending on the desired goal.

In this paper, the main goal is to detect bright components characterizing ICL. The discrimination operator $\mathcal{D}$ then becomes a way of classifying sources as ICL-type structures, denoted $\hat{\mathbf{x}}^{\rm{ICL}}$ and extract them from components associated to galaxies. A very simple way of doing so is to consider $j_{\rm{obj}}$ for each object and to use this parameter as a constraint since the characteristic size of galaxies is not the same as the characteristic size of any structural element of the ICL. One can then build in parallel of $\mathbf{z}$ the dictionary atom $\mathbf{z}^{\rm{ICL}}$ such as $\mathbf{z}^{\rm{ICL}}=\sum_{i}\hat{\mathbf{x}}_{(i)}^{\rm{ICL}}$. This atom is then added to the ICL synthesized dictionary $\mathbf{S}^{\rm{ICL}}$, and a fully restored ICL image can be computed by summing all its atoms. A discrimination based on the spatial position of the interscale maximum $k_{\rm{obj}}$ can also be applied, as one would also consider atoms describing galaxies belonging to a galaxy cluster (note that a catalog of the cluster member positions is needed in that case) for ICL fraction studies for example, or again to make sure that atoms associated with ICL are well centered on the galaxy cluster. More complete discrimination operators can be developed based on morphological properties of sources in the wavelet scales, like for instance granularity (i.e., the number of regions linked to an interscale maximum), depth of the interscale tree (i.e., the number of scales composing the tree), color of the restored object (when several bands are available), or others.


\begin{figure*}
  \centering
  \includegraphics[width=16cm]{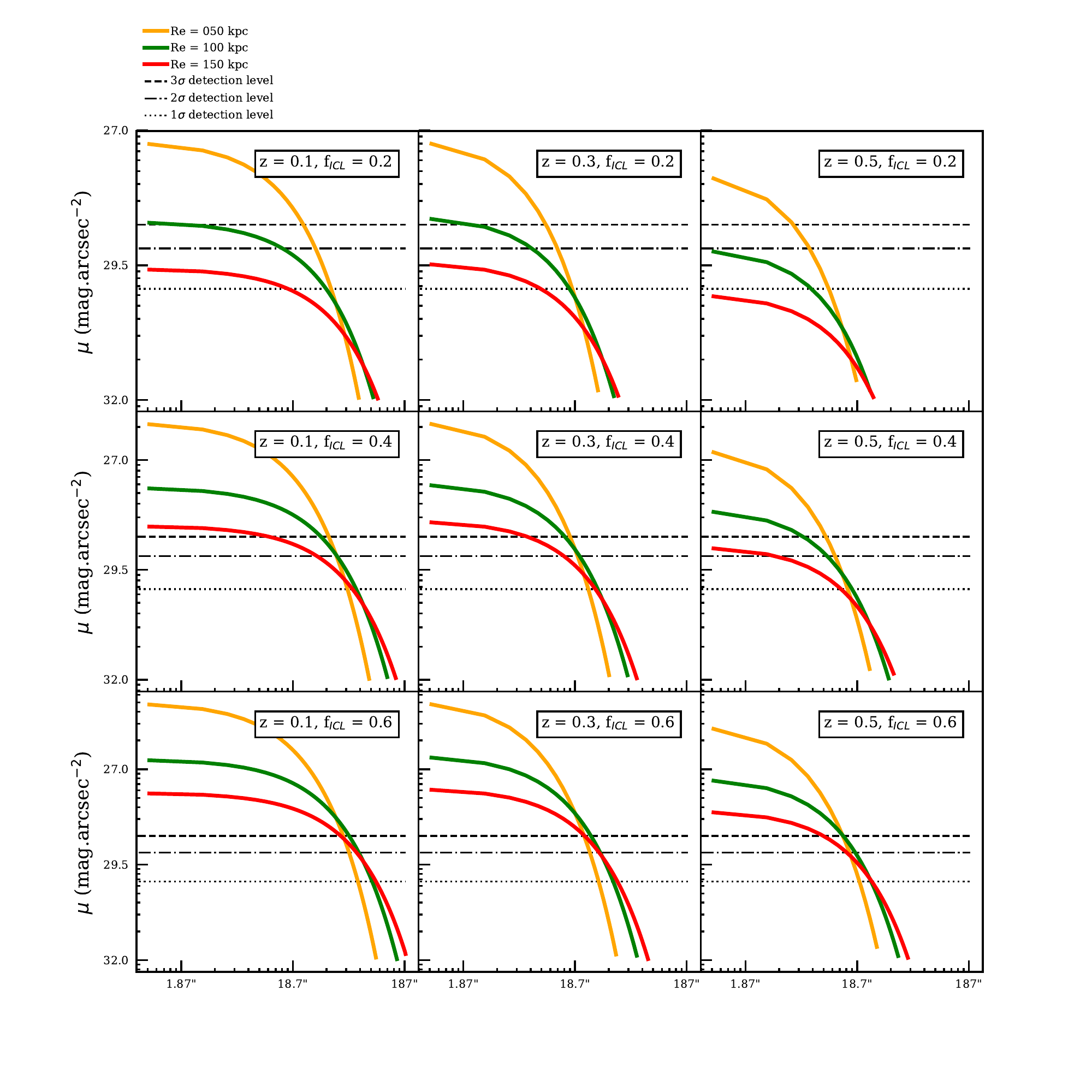}
  \caption{Evolution of integrated radial ICL profiles with input ICL fraction, redshift and half-light radius, computed from one of the simulated galaxy cluster catalogs. The typical $3\sigma$ detection limit (computed from an image containing only CCD noise) in our MegaCam type images is also displayed as a black dashed line. The half-light radius $r_{e}$ controls the concentration of the profile, and the redshift $z$ and the input ICL fraction $f_{\rm{ICL}}$ control its amplitude, with a stronger effect from $f_{\rm{ICL}}$. As the detection of ICL is strongly background limited, the outskirts (or the totality in some cases) of the profiles actually fall under the detection limit and are lost. This loss is estimated in Section~\ref{sec:application_of_detection_methods} (also see Figure~\ref{fig:icl_relative_error_clean}).}
  \label{fig:radial_profile}
\end{figure*}

\begin{figure*}
  \centering
  \includegraphics[width=16cm]{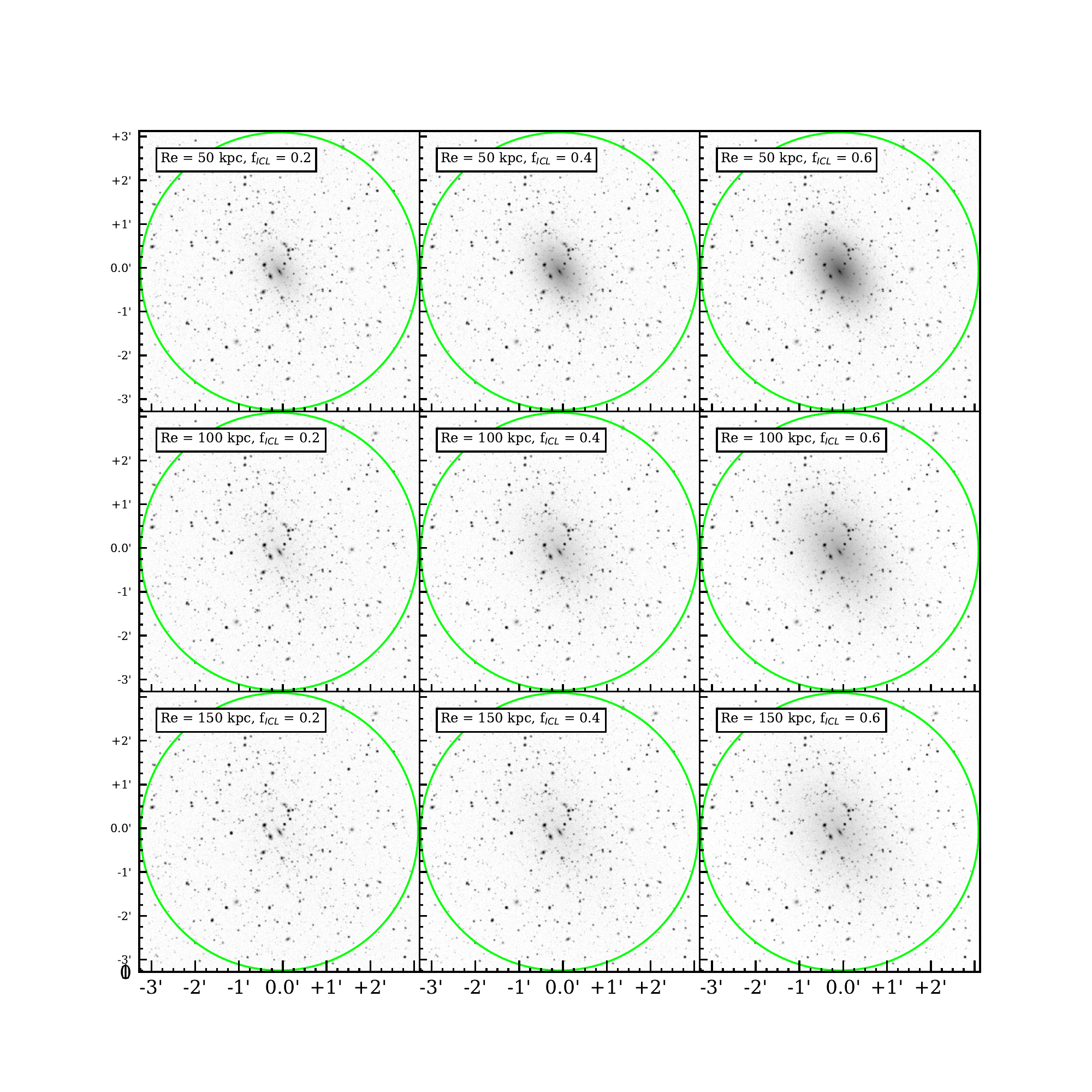}
  \caption{Mock MegaCam images of a simulated galaxy cluster at redshift $z=0.1$ with different ICL light profiles (the cluster is the same as that used for Figure~\ref{fig:radial_profile}). The concentration of the profile is controlled by the half-light radius $r_{e}$ while its amplitude is controlled by the input ICL fraction $f_{\rm{ICL}}$. The green circle is the radius $r_{\rm{in}}=350$ kpc, delimiting the core of the cluster. At this redshift, it covers almost the entire image. The intensity scale is logarithmic to highlight the faint ICL halo.}
  \label{fig:ellipticities}
\end{figure*}


\section{Simulations}
\label{sec:simulations}

A new detection algorithm such as \texttt{DAWIS} must be tested and compared to more traditional methods. For this purpose, we take an image analysis approach and create mono-chromatic mock images of galaxy clusters simulated with the \texttt{Galsim} package \citep{Rowe2015}, emulating the photometric aspect of galaxy clusters (galaxies + ICL) and the properties of the Canada France Hawaii Telescope (CFHT) wide field camera MegaCam. The choice of MegaCam has been made to prepare upcoming ground surveys such as the \textit{UNIONS} survey \citep{Ibata2017}. Note however that the same approach can (and should in the future) be applied to simulations of HST images for example. 

From an astrophysics point-of-view, this approach is not the most realistic since the simulated images contain only light profiles and an artificial background. However, it allows a complete control over the different components included, and allows us to compare different detection methods and their performances in different situations. Note that even if they are more realistic, it is not possible to do the same with $N$-body or hydrodynamical simulations since the way of defining ICL in those simulations is also subject to debate \citep{Rudick2011, Tang2018}.


\subsection{Photometric calibration}
\label{subsec:photometric_calibration}


We retrieved the MegaCam properties on the MegaPrime website\footnote{https://www.cfht.hawaii.edu/Instruments/Imaging/Megacam/}. The MegaCam-type images are set with a pixel scale $s = 0.187$ arcsec/pixel, a detector gain $g = 1.62$ e$^-$/ADU, an exposure time $t_{\rm{exp}} = 3600$ seconds, and a readout noise level $\sigma_{\rm{readout}}=5$ e$^-$. We calibrate the photometry for the $r$-band with a zero point $ZP=26.22$, and set the sky surface brightness $\mu_{\rm{sky}}$ to the corresponding average dark sky value which is $\mu_{\rm{sky}}=21.3$ mag~arcsec$^{-2}$. The sky background level in ADU/pix is then given by:

\begin{equation}
    F_{\rm{sky}}^{\rm{ADU/pix}} = \frac{s^{2}t_{\rm{exp}}}{g} 10^{\frac{ \mu_{\rm{sky}} - ZP }{- 2.5}}.
\end{equation}

The sky background in our images is simulated using the \texttt{Galsim} function 'CCDNoise(sky\_level =  $F_{\rm{sky}}^{\rm{ADU/pix}}$, gain = $g$, read\_noise = $\sigma_{\rm{readout}}$ )', which generates a spatially flat noise composed of a Gaussian readout noise plus a Poissonian noise.

In this study we neither try to model in a refined way the MegaCam PSF and the associated complex scattered light halos, nor try to measure the impact of its extended wings, and we do not include any spatial variation or anisotropy to it, therefore choosing a constant value for the seeing. We model the PSF by a Moffat profile using the \texttt{Galsim} function 'galsim.Moffat' with parameters $\beta=4.765$ and a half light radius of $0.7\arcsec$, which is a generic value for the MegaCam seeing.


\subsection{Creating catalogs}
\label{subsec:catalogs}
We first create a galaxy cluster catalog. Since the vast majority of ICL detection in the literature has been done at low redshift \citep[$z < 0.5$, with few exceptions such as][]{Adami2005, Burke2012, Guennou2012, Ko2018}, we pick three redshift values such as $z\in [0.1, 0.3, 0.5]$ in which 10 different galaxy clusters per bin are simulated. This choice of redshift values is made to study the effect of cosmological dimming on the ICL detection for each detection method. While the number of clusters per bin seems at first fairly low, different parameter spaces for each cluster ICL profiles are explored, multiplying the number of processed images to several hundreds and allowing us to lead a statistically significant study.

A NFW \citep{Navarro1997} Dark Matter (DM) gravitational potential is simulated at the center of an image for each cluster. The mass of the potential is set to an average value of $10^{15}$ M$_{\odot}$ and its concentration follows the N-body simulation concentrations of \citet{Klypin2016}. Recent works have shown that the spatial distribution of ICL should follow the concentration of the DM halo \citep{Montes2019a}. This is not the case in our simulations, and we do not explore the effects of the mass and concentration parameters on our results. Our goal is to mimic the photometric aspect of galaxy clusters and their ICL, and not to determine the physical properties of their gravitational potential or the impact of such parameters on ICL.

Galaxy catalogs for each cluster are simulated by drawing galaxy properties (redshift $z_{\rm{gal}}$, half-light radius $r_{h}$, apparent magnitudes in the $R_{\rm{C}}$ band $m_{R_{\rm{C}}}$, and the ellipticity parameters $\epsilon_{1}$ and $\epsilon_{2}$) from the COSMOS~2015 catalog \citep{Laigle2016} following a Poissonian distribution. First, a homogeneous field of galaxies is generated following the COSMOS~2015 field galaxy luminosity function (GLF) normalized by the size of the simulated image. Once the galaxy field is set, the cluster members are also drawn from the COSMOS~2015 catalog, but this time specifically following the GLF in the same redshift bin as the cluster redshift. The field GLF number counts are rescaled with the projected 2D mass density profile of the cluster, so that the number counts reflect those of the cluster rather than those of the field. By doing so, we assume a constant mass-to-light ratio in the cluster. Galaxy distances to the center are imposed by the cluster mass profile but their position angles are chosen randomly. 

These galaxy catalogs do not respect all the galaxy cluster properties such as the morphological segregation or the actual surface distribution of galaxies in a cluster (which is usually not exactly proportional to a NFW halo), but represent fairly well typical galaxy clusters at optical wavelengths: an overdensity of galaxies spatially following a halo mass profile superimposed on randomly placed field galaxies. Since the position or the fraction of elliptical and spiral galaxies in the cluster images should not impact in a significant way the detection of ICL, we do not add any supplementary properties to the catalogs. We do not include foreground stars in our simulations either, as we try to estimate the different ways of separating galaxies from ICL in galaxy clusters and not the impact of strong contamination sources such as these stars. Another missing component in these catalogs is a BCG at the center of the galaxy cluster, which is added by hand later (see Section~\ref{subsec:generating_GLP}).

We chose a size for the images and the galaxy catalogs of $383\times383\arcsec$, corresponding to images of $2048\times2048$ pixels for the CFHT/MegaCam-type images. This size is very convenient as it is possible to run \texttt{DAWIS} on a statistically significant sample of these images while giving exactly a limit of $J = 10$ on the number of wavelet scales (see Section~\ref{subsec:IUWT}). 

\subsection{Generating galaxy light profiles}
\label{subsec:generating_GLP}

The galaxy luminosity profiles are generated using the modeling package \texttt{Galsim}, in a way similar to \citet{Martinet2019}. Galaxies are represented by a S\'ersic profile and drawn into the images using the function 'galsim.Sersic'. The input half light radius $r_{h}$ is taken from the catalogs computed in Section~\ref{subsec:catalogs}, and the value of the S\'ersic index $n$ is drawn randomly following a uniform distribution with $n \in \left[ 0.5, 5.5\right]$. The input flux in ADU of each galaxy is computed with the relation

\begin{equation}
\label{eq:Fgal}
F_{\rm{gal}} = \frac{t_{\rm{exp}}}{g} 10^{\frac{m_{\rm{gal}}-\rm{ZP}}{-2.5}},
\end{equation}
\noindent
with $g$ being the detector gain, $t_{\rm{exp}}$ the exposure time, ZP the Zero Point of the image, and $m_{\rm{gal}}$ the galaxy magnitude from the catalogs computed in Section~\ref{subsec:catalogs}. The ellipticity of each galaxy is then computed with the function 'galsim.Shear($e_{1} = \epsilon_{1}$, $e_{2} = \epsilon_{2}$)'. Each galaxy profile is also convolved with the instrument PSF (see Section~\ref{subsec:photometric_calibration}) using the function 'galsim.Convolve'.

To compute the flux of a cluster, we retain only its galaxies within 350 kpc from its center. Applying a physical scale cut allows us to compare different redshift cases coherently. This ensures that there is no redshift bias when computing the total flux of a cluster, as low redshift galaxy clusters display an apparent size larger than high redshift ones. Real galaxy clusters are usually much larger of course, with sizes up to a few Mpc, but the measure of ICL fractions is dependent on the spatial extent which is probed and on the number of cluster members in it. We therefore impose this physical radius as a membership constraint, which is also chosen so the equivalent image radius of clusters in the redshift bin $z=0.1$ ($r\sim189.8$~\arcsec, the largest radius of the three redshift values) fits in CFHT MegaCam-type images of $2048\times2048$ pixels. The flux of a galaxy member is then added to the total flux of the galaxy cluster $F_{\rm{cluster}}$, that will then be used to compute the ICL luminosity profile.

A BCG is also simulated in the middle of each image by simulating a S\'ersic profile. The BCG flux $F_{\rm{BCG}}$ is computed by taking the flux of the brightest galaxy of the image and applying to it a cosmological dimming factor corresponding to the cluster redshift. This is done by multiplying its flux by a factor $(1+z)^{4}$ corresponding to the initial redshift value of the galaxy, and then dividing it by the same factor but with the cluster redshift value instead. In order to avoid cases where a very bright foreground galaxy is drawn into the image, giving a too high value for $F_{\rm{BCG}}$, we add the constraint that the absolute magnitude of the BCG is not brighter than $-23$. Following values from the literature \citep{Gonzalez2005, Seigar2007, Durret2019}, the BCG S\'ersic index is drawn from a uniform distribution in the conservative range $\left[1.5, 5.5\right]$, and $F_{\rm{BCG}}$ is added to the cluster total flux $F^{\rm{input}}_{\rm{cluster}}$. We assign to the BCG the same ellipticity as the one we apply to the ICL light profile generated in the next section. 

\subsection{Generating the ICL light profile}
\label{subsec:subsec:generating_ICL_light_profiles}
The procedure to simulate an ICL luminosity profile in a galaxy cluster is relatively unknown, with a variety of biases in the literature due to the various recipes applied to detect it in real images, and by the fact that ICL can show different morphologies both from cluster to cluster and when changing the observation wavelength. This is true for the smooth component of ICL, without even mentioning other sub-structures (tidal streams, shells...). Another unknown area is the fact that there is still no meaningful proof that the stellar populations emitting the ICL and the ones composing the galaxies can be disentangled in a consistent way without star kinematic information and using only photometric data. However for simplicity we make here the assumption that the ICL has its own light profile and we choose to simulate for each cluster a large exponential ICL profile with no sub-structures. While this may not be the most realistic approach, it allows us to probe at first order how the different detection methods we are testing act to disentangle a simple but faint and extended source superimposed on small bright objects. In addition, several studies \citep{Gonzalez2005, Seigar2007, Durret2019} found that the best fits for BCG+ICL systems were usually two-component fits with an internal S\'ersic profile and an external profile (exponential or S\'ersic depending on the study).

The ICL profile is simulated using the \texttt{Galsim} function 'galsim.Exponential(half\_light\_radius $=r_{e}$, flux $=F^{\rm{input}}_{\rm{ICL}}$)'. In order to obtain a profile which agrees with the cluster in the same image, we compute the ICL flux in ADU from the total cluster flux $F^{\rm{input}}_{\rm{cluster}}$ as

\begin{equation}
    F^{\rm{input}}_{\rm{ICL}} =  \frac{f^{\rm{input}}_{\rm{ICL}}.F^{\rm{input}}_{\rm{cluster}}}{1-f^{\rm{input}}_{\rm{ICL}}},
\end{equation}

\noindent with $f^{\rm{input}}_{\rm{ICL}}$ the input ICL fraction in our images. In the literature, the measured ICL fractions of galaxy clusters at intermediate redshifts usually range from 0.1 to 0.4 \citep{Montes2018, Jimenez-Teja2018}. Based on these values, on the facts that part of our simulated profile is masked by the sky background and that the measured ICL fractions will always be lower than the input ICL fractions in our images, we set $f^{\rm{input}}_{\rm{ICL}}\in\left[0.2, 0.4, 0.6\right]$.

While the amount of flux in the ICL profile is controlled by the input ICL fraction, another driver parameter for the profile average surface brightness is the half-light radius $r_{e}$, as it controls the concentration of the profile: while a very extended profile would likely fall under the sky background level, a profile with the same input ICL fraction but with smaller $r_{e}$ would stand out above. The measured half-light radii from the literature cover a very wide range as they strongly depend on the nature of the study and the quality of the applied fitting method, and one can find values of a few tens to several hundreds of kpc in the most extreme cases \citep{Gonzalez2005, Kluge2020, Durret2019}. In addition and as stated in Section~\ref{subsec:generating_GLP}, the characteristic size of our galaxy clusters is set by a physical radius of 350 kpc from the cluster center, giving an intrinsic upper limit in our simulations for our choice of values for $r_{e}$, as an ICL profile with a larger half-light radius would make no sense. This led us to set $r_{e}=\left[ 50, 100, 150 \right]$ kpc, a range of values probing quite concentrated ICL profiles (50 kpc) and extended ones (150 kpc).

As the effects of the cosmological dimming, the $r_{e}$ parameter and the $f^{\rm{input}}_{\rm{ICL}}$ parameter on the ICL surface brightness are degenerated, we show in Figure~\ref{fig:radial_profile} ICL surface brightness radial profiles with different sets of values for these parameters to illustrate their impact. We also display the typical $3\sigma$ detection limit in our images for a better visualisation (see Section~\ref{subsec:SBT} for details on how the detection limit is computed). We note that concentrated profiles ($r_{e}=50$ kpc) display in their inner part surface brightnesses brighter than $\mu=26.5$ mag~arcsec$^{-2}$ and could be qualified as 'unrealistic', but we nevertheless include them in the study. We also note that some ICL profiles fall completely under the detection limit, making it difficult for traditional detection methods to characterize them.

To summarize, we simulate ten galaxy clusters per redshift bin, and for each of them we vary the $r_{e}$ and $f^{\rm{input}}_{\rm{ICL}}$ parameters in three bins each, giving for each cluster nine different images corresponding to nine different ICL profiles. As we also want to ensure a diversity of morphologies for the ICL, we apply a shear to each elliptical profile with the function 'galsim.Shear($e = \epsilon$, beta$ = \theta$)' with $\epsilon$ the magnitude of the shear in the \texttt{Galsim} distortion definition drawn in the range $\left[0.0, 0.8 \right]$ and $\theta$ the angle of the shear drawn in the range $\left[ 0, 180 \right]$ degrees, both following a uniform distribution. For consistency, the same shear is applied to the nine different galaxy cluster ICL profiles (see Figure~\ref{fig:ellipticities} for an example showcasing the nine MegaCam images of one of the clusters at $z=0.1$).

\subsection{Drawing images}
\label{subsec:drawing_simulated_images}

We draw a full set of 270 MegaCam-type images using the parameters of Section~\ref{subsec:photometric_calibration}. In parallel to the generation of these generic cluster images (hereafter GAL+ICL+NOISE), we create alternative MegaCam-type images of these clusters: images containing only the galaxies and the noise (GAL+NOISE), and images containing only the ICL and noise (ICL+NOISE). These alternative versions will allow us to constrain the contamination due to the superposition of galaxies and ICL and the limitations of the different detection methods tested in Section~\ref{sec:application_of_detection_methods}. Note that another type of MegaCam-type image is generated during this step specifically for the profile fitting method (see Section~\ref{subsec:perfect_fitting} for details). In total, we analyze 270 different GAL+ICL+NOISE MegaCam-type images with a great diversity of ICL light profiles, both in morphology and in flux. We also analyse the equivalent 270 ICL+NOISE MegaCam-type images and 30 GAL+NOISE MegaCam-type images (as there are 30 different galaxy cluster catalogs), making a total of 540 images, which allow us to statistically evaluate the efficiency of the three detection methods we are testing.



\begin{figure*}
\centering
\includegraphics[width=16cm]{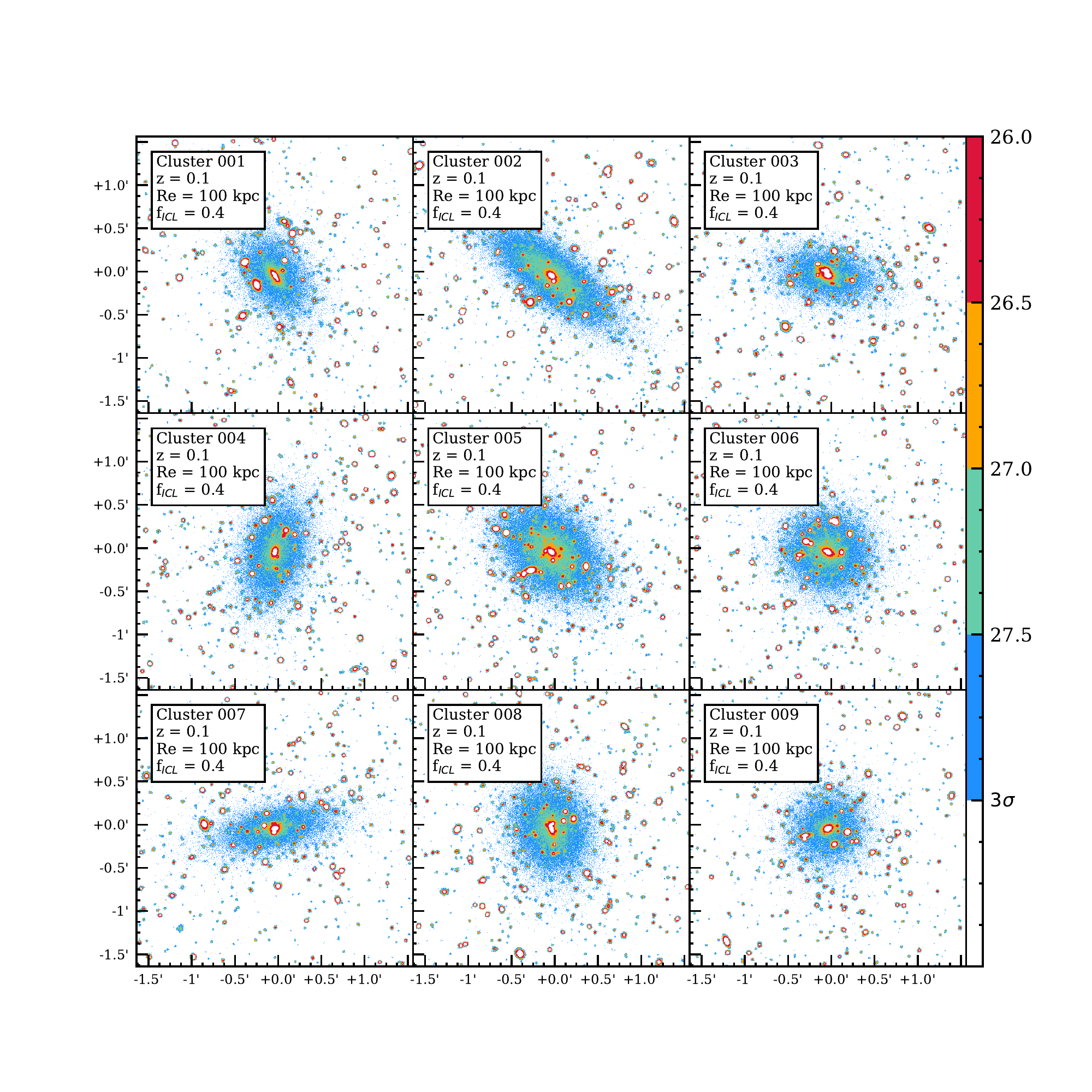}
\caption{GAL+ICL+NOISE MegaCam-type images of nine galaxy clusters with varying surface brightness thresholds (the color bar indicates the surface brightness threshold: $26\le\mu_{\rm{SBT}}\le28$ mag~arcsec$^{-2}$). We display here clusters with generic ICL profiles (half-light radius of $r_{e} = 100$ kpc, input ICL fraction of $f_{\rm{ICL}}=0.4$ and a redshift $z=0.1$) with different ellipticities. The pixels with $\mu<26$ mag~arcsec$^{-2}$ as well as the non-significant pixels under the detection limit are masked (e.g., white). The galaxy S\'ersic profile outskirts contaminating the detection of ICL are easily discernible by eye everywhere in the image when the threshold used to mask galaxies is too bright. In the same way, part of the ICL profile is accidentally masked with galaxies when the threshold used is too faint.}
\label{fig:SBT}
\end{figure*}




\section{Applications of detection methods}
\label{sec:application_of_detection_methods}

In this section we describe and apply three different methods to detect ICL in the simulated images of galaxy clusters from Section~\ref{sec:simulations}. The methods are labeled as following: surface brightness threshold (SBT), ideal profile fitting (PF), and \texttt{DAWIS}. We apply the following procedure for each detection method:

\begin{enumerate}

    \item Application of the method to the ICL+NOISE MegaCam-type images, to measure separately the ICL flux, hereafter $F^{\rm{sep}}_{\rm{ICL}}$.
    \item Application of the method to the GAL+NOISE MegaCam-type images, to measure separately the flux of cluster galaxies, hereafter $F^{\rm{sep}}_{\rm{cluster}}$.
    \item Computation of the ICL fraction $f^{\rm{sep}}_{\rm{ICL}} = F^{\rm{sep}}_{\rm{ICL}} / (F^{\rm{sep}}_{\rm{ICL}} + F^{\rm{sep}}_{\rm{cluster}}) $. Comparison of the results with input values to constrain the method limitations.
    \item Application of the method to the GAL+ICL+NOISE MegaCam-type images, to measure the fluxes $F_{\rm{ICL}}$ and $F_{\rm{cluster}}$.
    \item Computation of $f_{\rm{ICL}} = F_{\rm{ICL}} / ( F_{\rm{ICL}} + F_{\rm{cluster}} )$. Comparison of the results with $F^{\rm{sep}}_{\rm{ICL}}$, $F^{\rm{sep}}_{\rm{cluster}}$ and $f^{\rm{sep}}_{\rm{ICL}}$ to constrain the effect of the superposition of galaxies and ICL.
    
\end{enumerate}

\subsection{Surface brightness threshold}
\label{subsec:SBT}

We make the choice to consider a threshold  $\mu_{\rm{SBT}} = 26.5$ mag~arcsec$^{-2}$ on our images to constrain this approach and compare it to other detection methods, as this value is still the most frequently used in the literature (we also try different threshold values as displayed in Appendix~\ref{sec:varying_sbt}, but only describe in depth the case \sbt). The procedure is straightforward: for the GAL+ICL+NOISE image, pixels with values brighter than this SBT (taking into account cosmological dimming for the threshold) are masked (their pixel values are set to zero). The pixels included in $r_{\rm{in}}$ and with values greater than the $3\sigma$ detection limit (computed by applying a $\sigma$-clipping algorithm on an image containing only the CCD noise generated in Section~\ref{subsec:photometric_calibration}) are defined as significant and are associated to ICL. $F_{\rm{ICL}}$ is computed by summing their values. The pixels that have values brighter than $\mu_{\rm{SBT}}$ are grouped into regions of pixels. A cross-correlation between the catalog of cluster member positions and the position of the pixel with the highest value of each newly defined region of pixels is performed. Matching regions are identified as cluster members and the fluxes of the cluster galaxies $F_{\rm{cluster}}$ are computed by summing their values.

This method has a few intrinsic limitations. First, part of the ICL light profile is lost under the $3\sigma$ detection limit in the original image, and another part is lost as it is masked by galaxies. The detected ICL flux is also contaminated by galaxy flux, both from the outskirts of galaxies falling under $\mu_{\rm{SBT}}$ (see Figure~\ref{fig:SBT} where a quick eye-test showcases this effect) and from undetected galaxies under the $3\sigma$ detection limit. To constrain these effects, a similar procedure is applied to the images with separated galaxies and ICL. $F^{\rm{sep}}_{\rm{cluster}}$ is computed from pixels with values brighter than $\mu_{\rm{SBT}}$ in the GAL+NOISE image. As we want to constrain the loss of ICL flux masked by galaxies, the same masks are applied to the ICL+NOISE image (therefore masking part of the ICL light profile). After that, if there are any  pixels left brighter than $\mu_{\rm{SBT}}$ in the ICL+NOISE image, we also mask them, as part of the ICL profile might be brighter than this threshold and also masked with galaxies. $F^{\rm{sep}}_{\rm{ICL}}$ is then computed from the significant pixel values above the $3\sigma$ detection limit. Therefore the difference between $F^{\rm{input}}_{\rm{ICL}}$ and $F^{\rm{sep}}_{\rm{ICL}}$ gives the effect of the masks and of the background noise on the quantity of ICL detected, while the difference between $F^{\rm{sep}}_{\rm{ICL}}$ and $F_{\rm{ICL}}$ gives the effect of galaxy contamination.

We emphasize the fact that in the literature, studies using this method to separate ICL from galaxies in observational data apply a first pass to extensively mask foreground objects and avoid as much as possible contamination from bright sources before applying the SBT \citep[][for two examples]{Burke2015, Montes2018}, using for example the \texttt{SExtractor} package. We do not apply such a first pass here. On the first hand, applying this method could lead to larger contamination effects in the detection of ICL in our simulations. On the other hand, relatively to the quantity of actual ICL flux in the image, more ICL should actually be detected in our work since the spatial area degraded by masks is smaller.

\subsection{Profile fitting}
\label{subsec:perfect_fitting}
The second detection method is based on the galaxy intensity profile fitting, where the fitted profiles are removed from the image before detecting ICL in the residuals. The result of such a method strongly depends on the quality of the fits and on the quality of the analysis package used to perform the analysis \citep[one could cite performant packages such as \texttt{Galfit} or \texttt{CICLE};][]{Peng2002, Jimenez-Teja2018}. Here again there is a large variety of studies which are different from one another, and we do not try to exactly replicate each of them. Our approach here is to consider an ideal profile fitting (PF) method, where the fitting algorithm allows to perfectly recover the intensity profile of each detected galaxy. While it is unlikely that such an algorithm exists or will ever exist, it gives an upper limit to the quality of fitting methods and allows us to compare them easily with other approaches.

The procedure for this method is different from the other two, as we do not use exactly the same GAL+ICL+NOISE image. Instead, in parallel of the drawing of the galaxy light profiles in Section~\ref{subsec:generating_GLP}, we apply to each galaxy drawn the $3\sigma$ detection limit (the same as in Section~\ref{subsec:SBT}), and galaxies without any pixel above this limit are defined as undetected. Every undetected galaxy is added on both the generic GAL+ICL+NOISE image (used for the two other detection methods) and on a second image where we only draw these faint galaxies. The ICL profile and the noise (see Section~\ref{subsec:subsec:generating_ICL_light_profiles}) are then added to this image too. This image has the same properties as the generic GAL+ICL+NOISE image but without the light profile of detected galaxies, as if a fitting algorithm was able to perfectly remove them. The ICL flux $F_{\rm{ICL}}$ for this method is detected in this image by applying to it the detection limit and keeping the significant pixels included in $r_{\rm{in}}$. The cluster galaxy flux $F_{\rm{cluster}}$ is computed directly from the light profiles that are not drawn onto this image. Note that with this method, the detected cluster flux is the same for the GAL+NOISE image as for the GAL+ICL+NOISE image, \fluxgal  = \fluxgalsep

Even if the light profiles of galaxies are perfectly removed from the image, this method is still sensitive to galaxy contamination from undetected galaxies that are below the $3\sigma$ detection limit. It is also sensitive to the loss of ICL flux under the $3\sigma$ detection limit. To constrain these effects, we compute $F^{\rm{sep}}_{\rm{ICL}}$ by summing the values of significant pixels above the $3\sigma$ detection limit in the ICL+NOISE image. The difference between $F^{\rm{input}}_{\rm{ICL}}$ and $F^{\rm{sep}}_{\rm{ICL}}$ gives the amount of flux lost under the detection limit, and can be contrasted with the value from the SBT method which gives the amount of flux lost under the detection limit and behind the galaxy masks. The difference between $F^{\rm{sep}}_{\rm{ICL}}$ and $F_{\rm{ICL}}$ then gives the amount of ICL flux contaminated by faint undetected galaxies with this method, which can be once again contrasted with the values from the SBT method, where both unmasked outskirts of galaxies and undetected faint galaxies contaminate $F_{\rm{ICL}}$. We must also remember that a real profile fitting method would have ICL flux contaminated by residuals, and that we do not consider this effect here.

\subsection{\texttt{DAWIS} tuned parameters}
\label{subsec:dawis_tuned}

The detailed operating structure of \texttt{DAWIS} was already explained in Section~\ref{sec:dawis}, and we only give here some additional information about a few parameters. Various values for the \texttt{CLEAN} factor $\delta$ and the relative threshold parameter $\tau$ were tested for the \texttt{MORESANE} algorithm and \citet[][]{Dabbech2015} suggested values $\tau \in \left[ 0.8, 0.9\right]$ and $\delta \in \left[0.1, 0.2\right]$ for a good convergence of their algorithm in parallel of a good quality of source restoration. However, their tests were run on more complicated simulations than our mock images of galaxy clusters; the fact that all the sources in our mock images are quasi circular and the absence of unreasonably bright objects such as foreground MW stars simplify the analysis in our case. We set $\tau = 0.9$ and $\delta=0.5$, a higher value which boosts a little the convergence of \texttt{DAWIS}. The same values for these parameters are used for all the processed images.

\begin{sidewaysfigure*}
    \includegraphics[width=\textwidth]{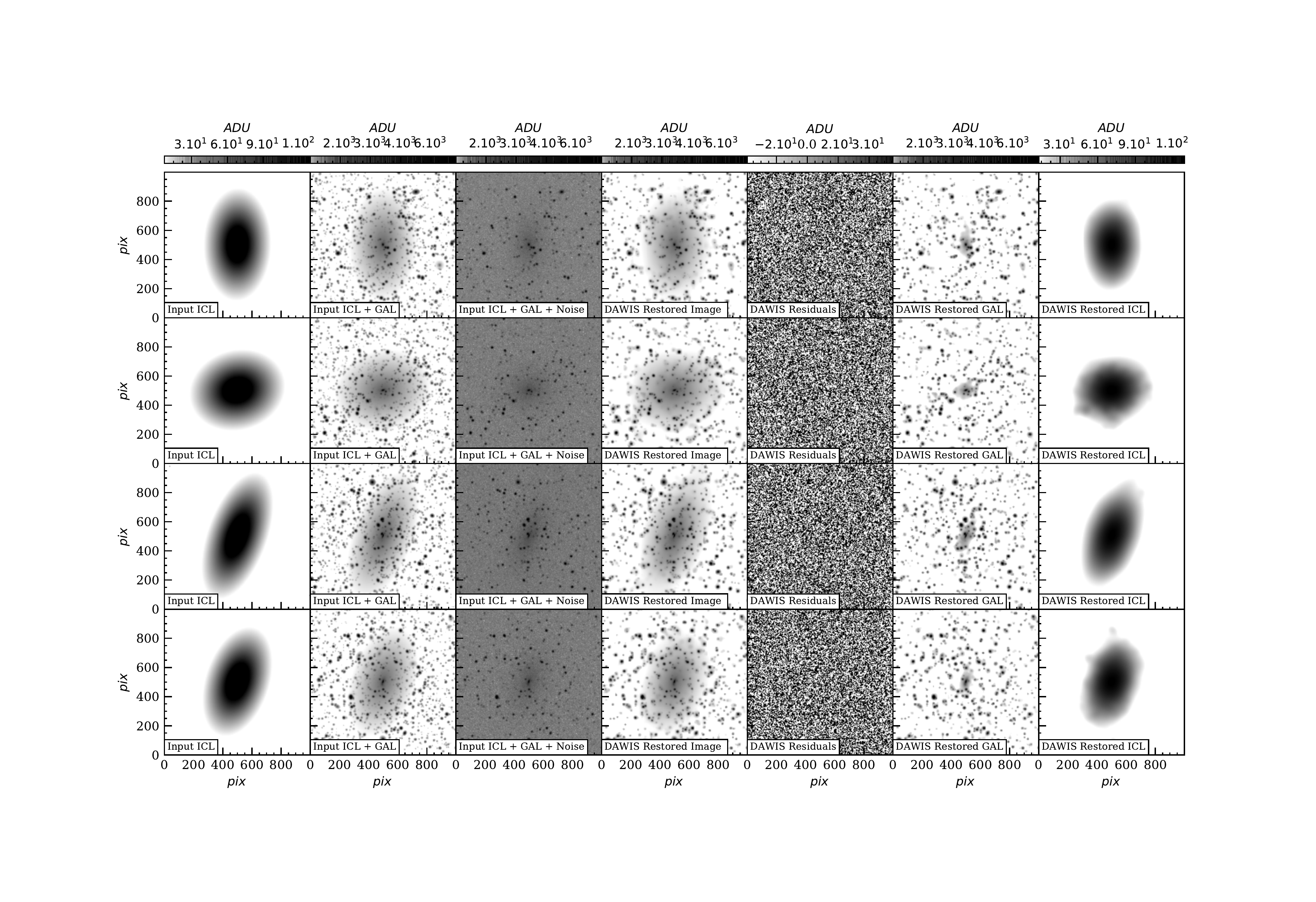}
    \caption{Sample of a few galaxy clusters processed by \texttt{DAWIS}. The clusters are at $z=0.3$, and the input ICL profile parameters are $r_{e}=100$~kpc and $f^{\rm{input}}_{\rm{ICL}} = 0.4$. In the first column are shown the input ICL profiles. The second column shows the cluster images with both ICL and galaxies but without noise. The third column shows the cluster images, with noise added, called GAL+ICL+NOISE images in the text. The fourth column shows the total denoised image restored by \texttt{DAWIS}. The fifth column shows the residuals image, which is the difference between the GAL+ICL+NOISE image and the \texttt{DAWIS} restored image. The sixth and seventh columns show the restored images containing atoms classified respectively as galaxies and ICL (see details in Section~\ref{subsec:dawis_tuned}). Note that the intensity scale is the same for columns 2, 3, 4 and 6, but that the contrast is greatly enhanced for columns 1, 5 and 7.}
    \label{fig:dawis_sample}
\end{sidewaysfigure*}


\begin{figure*}[h!]
\centering
\includegraphics[width=16cm]{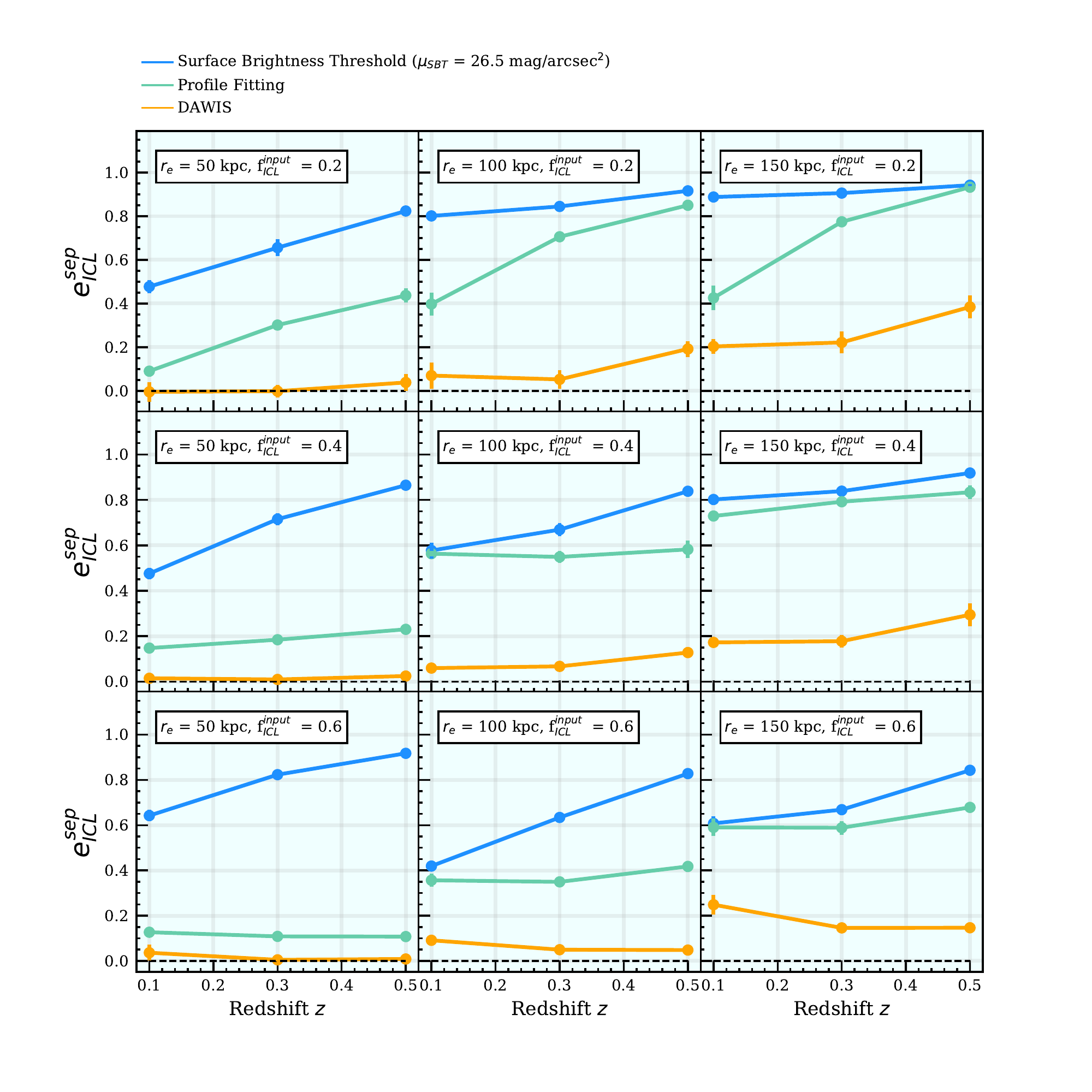}
\caption{Relative biases $e^{\rm{sep}}_{\rm{ICL}}$ (see Equation~\ref{eq:relative_error_icl_sep} in the text) given by the different detection methods and displayed against the three different ICL profile parameters (half-light radius $r_{e}$, input ICL fraction $f^{\rm{input}}_{\rm{ICL}}$ and redshift $z$). The biases have been computed for each of the 270 ICL+NOISE MegaCam-type images, and the displayed values correspond to the average of the ten images with the same input parameters. The error bars correspond to standard deviations across ten different clusters. Here the values for the PF method give the amount of ICL flux lost under the \DL (also see Figure~\ref{fig:radial_profile}).}
\label{fig:icl_relative_error_clean}
\end{figure*}

In order to measure ICL fractions in the GAL+ICL+NOISE images with \texttt{DAWIS}, one needs to identify the objects related to ICL and those related to cluster members, and to compute the associated synthesis atoms (see Section~\ref{subsec:atom_classification}). At each main iteration, each object going through the restoration step goes through a two-step selection. First, a cross-correlation between the interscale maximum spatial position of the object and the cluster galaxy catalog is performed. Then, if the wavelet scale of the interscale maximum $j_{\rm{obj}}$ is greater or equal to a value $j_{\rm{sep}}$ (corresponding to objects with characteristic sizes of at least $2^{j_{\rm{sep}}}$ pixels; we set $j_{\rm{sep}}=5$ for the GAL+ICL+NOISE MegaCam-type images) it is associated to ICL, otherwise it is associated to the cluster galaxies. In this way, we compute in parallel of the total restored image $z_{\rm{tot}}$ (containing the ICL and all galaxies) the total restored image of cluster members and the total restored image of ICL. A sample of galaxy clusters is shown in Figure~\ref{fig:dawis_sample} to illustrate the operating mode of \texttt{DAWIS}. $F_{\rm{ICL}}$ and $F_{\rm{cluster}}$ are given by the sums of the pixels of their respective images.

The main limitation of this approach comes from this very rough classification based on characteristic size. To constrain this limitation, we also run \texttt{DAWIS} without any classification on the ICL+NOISE image to measure $F^{\rm{sep}}_{\rm{ICL}}$ and on the GAL+NOISE image to measure $F^{\rm{sep}}_{\rm{cluster}}$. The first one can be contrasted with the values from the two other methods to showcase the efficiency of \texttt{DAWIS} to detect faint sources, as it is not limited in the same way by the $3\sigma$ detection limit in the original image. Then the difference between $F^{\rm{sep}}_{\rm{ICL}}$ and $F_{\rm{ICL}}$, as well as the one between $F^{\rm{sep}}_{\rm{cluster}}$ and $F_{\rm{cluster}}$ shows the effect of the atom classification.

\subsection{ICL and galaxies in separated images}
\label{subsec:results_separated}

We first analyse the results of each method on the ICL+NOISE and GAL+NOISE images. We compute for each cluster and for each method the relative biases:

\begin{equation}
    \label{eq:relative_error_icl_sep}
    e^{\rm{sep}}_{\rm{ICL}} = \frac{F^{\rm{input}}_{\rm{ICL}}-F^{\rm{sep}}_{\rm{ICL} }}{F^{\rm{input}}_{\rm{ICL}}} 
\end{equation}

\noindent
and

\begin{equation}
    \label{eq:relative_error_gal_sep}
    e^{\rm{sep}}_{\rm{cluster}} = \frac{F^{\rm{input}}_{\rm{cluster}}- F^{\rm{sep}}_{\rm{cluster}}}{F^{\rm{input}}_{\rm{cluster}}}
\end{equation}

\noindent
with $F^{\rm{sep}}_{\rm{cluster}}$ the flux of the cluster galaxies measured in the GAL+NOISE image and $F^{\rm{sep}}_{\rm{ICL}}$ the flux of ICL detected in the ICL+NOISE image.

\begin{figure}
\centering
\includegraphics[width=\hsize]{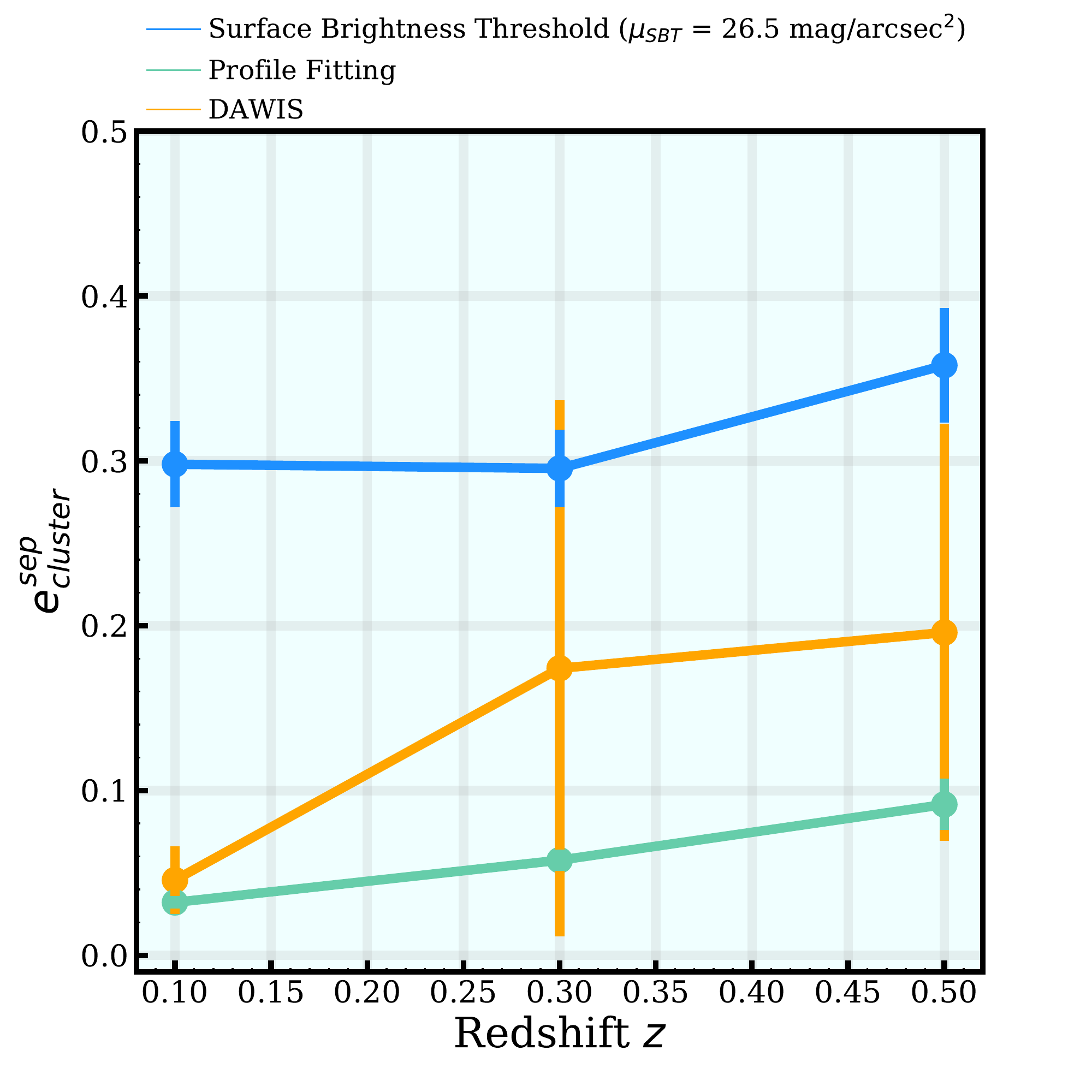}
\caption{Relative biases $e^{\rm{sep}}_{\rm{cluster}}$ (see Equation~\ref{eq:relative_error_gal_sep} in the text) given by the different detection methods and displayed against redshift $z$. They have been computed for each of the 30 GAL+NOISE MegaCam-type images, and the displayed values correspond to the average of the ten images at the same redshift. The error bars correspond to standard deviations across ten different clusters.}
\label{fig:gal_relative_error_clean}
\end{figure}


\begin{figure*}
\centering
\includegraphics[width=16cm]{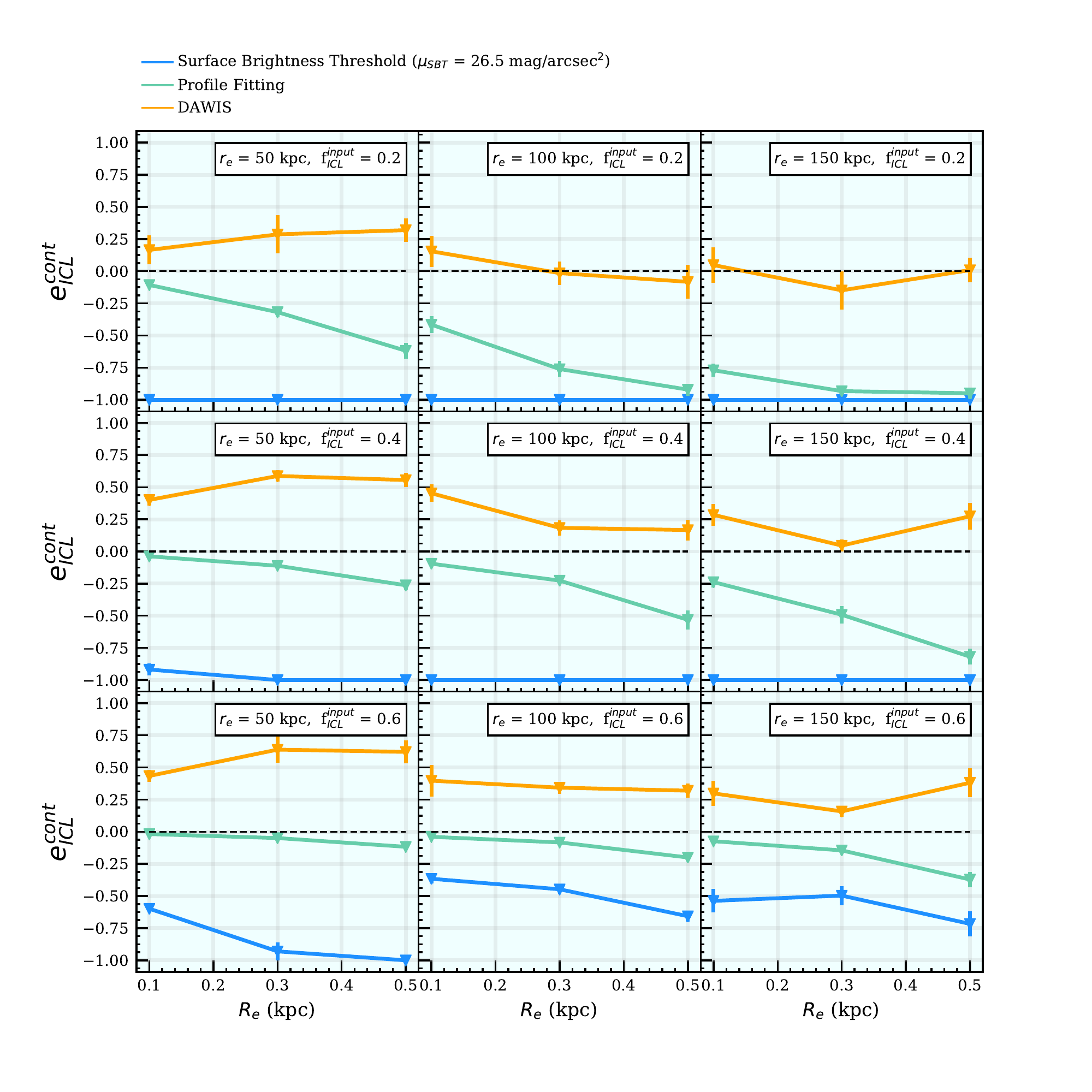}
\caption{Relative biases $e^{\rm{cont}}_{\rm{ICL}}$ (see Equation~\ref{eq:relative_error_icl_sep} in the text) given by the different detection methods and displayed against the three different ICL profile parameters (half-light radius $r_{e}$, input ICL fraction $f^{\rm{input}}_{\rm{ICL}}$ and redshift $z$). They have been computed for each of the 270 GAL+ICL+NOISE MegaCam-type images, and the displayed values correspond to the average of the ten images with the same input parameters. The error bars correspond to standard deviations across ten different clusters. The values of minus unity for the SBT method have actually been truncated for representation purposes (see text for more details).}
\label{fig:icl_relative_error_contamination}
\end{figure*}

\begin{figure*}
\centering
\includegraphics[width=16cm]{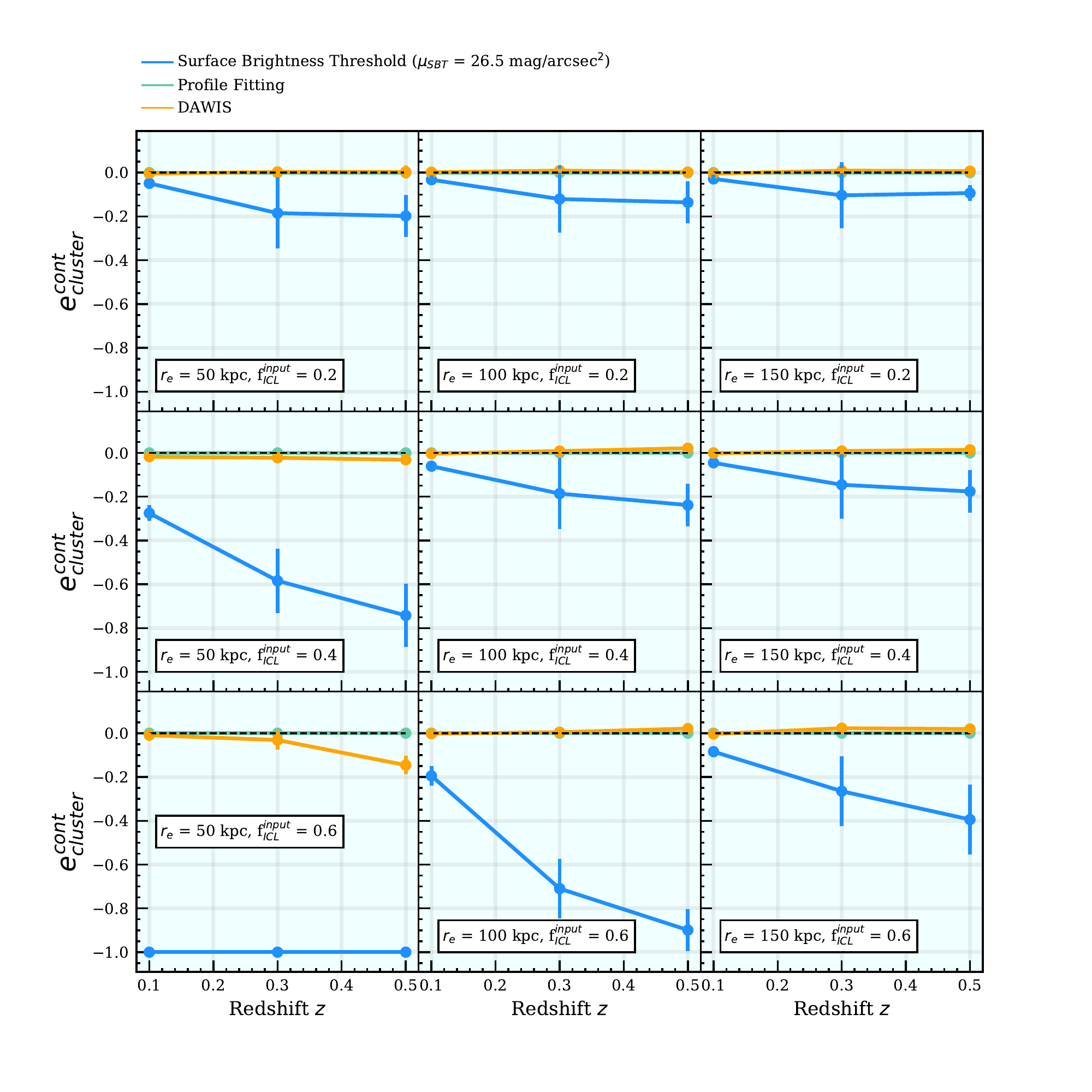}
\caption{Relative biases $e^{\rm{cont}}_{\rm{cluster}}$ (see Equation~\ref{eq:relative_error_gal_cont} in the text) given by the different detection methods and displayed against redshift $z$. They have been computed for each of the 270 GAL+ICL+NOISE MegaCam-type images, and the displayed values correspond to the average of the ten images at the same redshift. The error bars correspond to standard deviations across ten different clusters.}
\label{fig:gal_relative_error_contamination}
\end{figure*}


\begin{figure*}
\centering
\includegraphics[width=16cm]{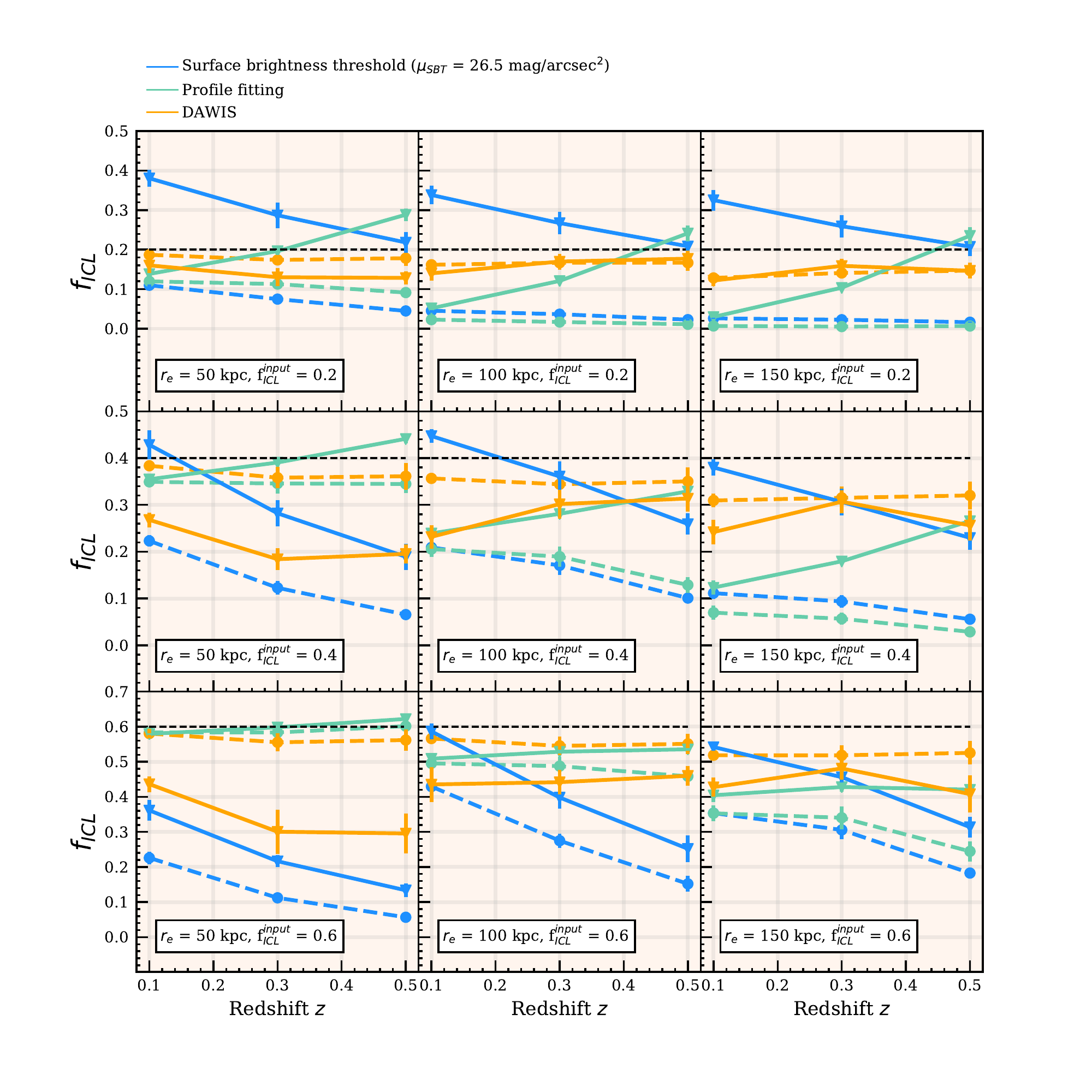}
\caption{ICL fractions displayed for the different ICL profile parameters (half-light radius $r_{e}$, input ICL fraction $f^{\rm{input}}_{\rm{ICL}}$ and redshift $z$) as measured by the different detection methods. The triangles with continuous lines are the ICL fractions \Ficl computed from each of the 270 GAL+ICL+NOISE MegaCam-type images, and the displayed values correspond to the average of the ten images with the same input parameters. The filled circles with dashed lines are the ICL fraction \Ficlsep computed from the images with separated galaxies and ICL. The error bars correspond to standard deviations across ten different clusters. The input ICL fraction is also displayed as a black horizontal dashed line for better visualization.}
\label{fig:ficl_all_group}
\end{figure*}
The values of $e^{\rm{sep}}_{\rm{ICL}}$ for the different detection methods are displayed in Figure~\ref{fig:icl_relative_error_clean} as a function of the various ICL light profile parameters. We first address the results for the PF method, which show in this particular case the amount of ICL flux lost under the $3\sigma$ detection limit (also see Figure~\ref{fig:radial_profile}). As expected, the loss of flux is minimal for concentrated and bright ICL profiles. The brightest the ICL profile, the more it stands out above the sky background. This is best shown by the case with $r_{e} = 50$ kpc and $f^{\rm{input}}_{\rm{ICL}} = 0.6$, where the values of $e^{\rm{sep}}_{\rm{ICL}}$ are lower than $\sim$0.2 at all redshifts for the PF method. This means that less than 20\% of the input ICL flux injected in the ICL profile is lost under the $3\sigma$ detection limit in that case.

For fainter and more extended profiles, the portion of ICL flux hidden under the \DL is higher. The relative bias  \eSICL for the PF method reaches strong values such as $\sim$0.9 for the faintest and most extended profiles ($z=0.5$, $r_{e}=150$~kpc and $f^{\rm{input}}_{\rm{ICL}} = 0.2$), meaning that around 90\% of the input flux injected in the ICL profile is not detected in this case.

In comparison to the PF method, the \eSICL values for the SBT method are always higher. This means that the SBT method detects consistently less ICL flux than the PF method. This is expected since it is not only background limited, but also features a second source of degradation of the ICL profile due to the masks. Note that this difference is really amplified for concentrated and bright profiles ($r_{e} = 50$ kpc and $f^{\rm{input}}_{\rm{ICL}} = 0.6$). This is due to the fact that part of the ICL profile is actually brighter than the SBT threshold \sbt\ (see Figure~\ref{fig:radial_profile}), and therefore is masked.

Of all three methods, \texttt{DAWIS} performs the best, as in most cases $e^{\rm{sep}}_{\rm{ICL}}$ is lower than $\sim$0.2, with an outlier at 0.4 for the most extended and faintest profile ($z=0.5$, $r_{e}=150$~kpc and $f^{\rm{input}}_{\rm{ICL}} = 0.2$). This clearly shows the efficiency of the wavelet based analysis to detect faint and extended sources, even if they mostly lie close or under the usual $3\sigma$ detection limit in the original image.

Figure~\ref{fig:gal_relative_error_clean} displays the values of the relative bias $e^{\rm{sep}}_{\rm{cluster}}$. For all methods $e^{\rm{sep}}_{\rm{cluster}}$ increases with redshift. This is due to cluster galaxies becoming fainter, diminishing the number of detections and resulting in a loss of detected flux. As expected, the best method here is the PF method, as the flux of the cluster galaxies $F^{\rm{sep}}_{\rm{cluster}}$ is computed directly from the galaxy profiles that are not drawn on the image (see Section~\ref{subsec:perfect_fitting} for more details). The bias in this case only comes from the faint undetected galaxies below the detection limit. This effect is rather weak as $e^{\rm{sep}}_{\rm{cluster}}$ for the PF method is lower than 0.1, meaning that this effect accounts for a loss of less than 10\% of the input flux of the cluster galaxies.

The second best performing method in average is \texttt{DAWIS} with mean \eSgal values lower than 0.2 at all redshifts. However, the error bars show discrepancies between different clusters at the same redshift. The less performing method is the SBT method with values around $\sim$~0.3 or higher for all redshifts. This is also expected as the Holmberg radius defining the threshold $\mu_{\rm{SBT}}=26.5$mag~arcsec$^{-2}$ is only a rough estimation of the size of galaxies. This method misses the outskirts of the S\'ersic profiles, in addition to flux from faint undetected galaxies.

When comparing Figure~\ref{fig:icl_relative_error_clean} and Figure~\ref{fig:gal_relative_error_clean}, we can see that the relative bias on ICL \eSICL is on average higher than the relative bias on the cluster galaxies \eSgal for the PF and SBT methods. At $z=0.5$ for example, both methods disclose \eSgal values lower than 0.4. For faint and extended profiles at the same redshift ($z=0.5$, $r_{e}=150$~kpc and $f^{\rm{input}}_{\rm{ICL}} = 0.2$), the \eSICL are twice this value. This shows that these standard detection methods are more limited by the detection of the ICL profile than by the detection of galaxies. On the contrary, \texttt{DAWIS} is more stable and discloses equivalent values lower than 0.2 at all redshifts for both \eSICL and \eSgal.

The resulting ICL fractions $f^{\rm{sep}}_{\rm{ICL}}$ for separated galaxies and ICL are displayed in Figure~\ref{fig:ficl_all_group} with dashed lines. Comforting our previous analyses of $e^{\rm{sep}}_{\rm{ICL}}$ and $e^{\rm{sep}}_{\rm{cluster}}$, in average \texttt{DAWIS} measures $f^{\rm{sep}}_{\rm{ICL}}$
better than the two other detection methods. For bright and concentrated profiles ($r_{e}=50$~kpc and $f^{\rm{input}}_{\rm{ICL}} = 0.6$), the loss of ICL flux due to masking with the SBT method is very important and results in an large underestimation of \Ficlsep. This sets the SBT as the worst method to detect ICL in this case. On the other hand, the PF method obtains excellent results for the same ICL profiles and performs even better than \texttt{DAWIS}. For faint and extended ICL profiles ($r_{e}=150$~kpc and $f^{\rm{input}}_{\rm{ICL}} = 0.2$), both the SBT and PF methods converge to the same results and underestimate greatly the ICL fraction \Ficlsep.

However, while the results of the SBT and PF methods look similar for these faint and extended ICL profiles, Figures~\ref{fig:icl_relative_error_clean} and \ref{fig:gal_relative_error_clean} showed that the measures of $F^{\rm{sep}}_{\rm{ICL}}$ and $F^{\rm{sep}}_{\rm{cluster}}$ for both methods are different. The SBT method detects less ICL flux than the PF method, which is compensated by the fact that it also detects less galaxy flux. This illustrates for the first time in this work how a bias on the measure of ICL flux can compensate a bias on the measure of cluster galaxy flux when computing ICL fractions. An amplified version of this effect will be showcased in the next section, where the contamination effects due to the superposition of galaxies and ICL are discussed.

\subsection{Results on superimposed galaxies and ICL}
\label{subsec:results_superimposed}

We now show the results of the three detection methods on the GAL+ICL+NOISE MegaCam-type images. We measure both $F_{\rm{ICL}}$ and $F_{\rm{cluster}}$ in these images, and compute the following relative biases

\begin{equation}
    \label{eq:relative_error_icl_cont}
    e^{\rm{cont}}_{\rm{ICL}} = \frac{F^{\rm{sep}}_{\rm{ICL}}- F_{\rm{ICL}}}{F^{\rm{sep}}_{\rm{ICL}}} 
\end{equation}

\noindent
and

\begin{equation}
    \label{eq:relative_error_gal_cont}
    e^{\rm{cont}}_{\rm{cluster}} = \frac{F^{\rm{sep}}_{\rm{cluster}}- F_{\rm{cluster}}}{F^{\rm{sep}}_{\rm{cluster}}}.
\end{equation}

\noindent
giving us a measure of the effect of the superposition of galaxies on ICL. We call it the contamination effect.
The values of $e^{\rm{cont}}_{\rm{ICL}}$ are displayed in Figure~\ref{fig:icl_relative_error_contamination}. We first consider the values of $e^{\rm{cont}}_{\rm{ICL}}$ for the PF method. All the values are negative, meaning that more flux is detected and associated to ICL in the GAL+ICL+NOISE image than in the ICL+NOISE image. This shows the contamination effect by faint undetected galaxies, as it is the only difference between the measure of $F^{\rm{sep}}_{\rm{ICL}}$ and $F_{\rm{ICL}}$. This effect is increasing with redshift, as a higher and higher number of cluster galaxies fall under the \DL. For faint and extended ICL profiles \faint, this contamination effect becomes prominent. The value of \eCgal reaches values of almost minus unity, meaning that in this case $F_{\rm{ICL}}$ is twice as high as $F^{\rm{sep}}_{\rm{ICL}}$ and that the undetected galaxies account for half the detected flux associated to ICL.
In comparison to the PF method, the values of \eCICL for the SBT method show the effect of galaxy outskirt contamination (see Figure~\ref{fig:SBT} for visualisation) in addition to the contamination effect of faint undetected galaxies. The values are also always negative, meaning that with this method more flux is detected and associated to ICL in the GAL+ICL+NOISE image than in the ICL+NOISE image. The first problem to address is the shockingly large bias values of this method. We had to truncate them to fit them in the plots of Figure~\ref{fig:icl_relative_error_contamination}, as \eCICL reaches values of $-10$ for faint and extended profiles \faint. This means that the values of \fluxicl\ are ten times higher than the values of \fluxiclsep. This shows the very strong contamination effect of galaxy outskirts in this case, which completely dominates the detected flux associated to ICL with the SBT method. The detection of ICL in these cases can be qualified as blatantly wrong.

\texttt{DAWIS} discloses disparate values for $e^{\rm{cont}}_{\rm{ICL}}$. In average the bias is positive, showing that less ICL flux is detected in the GAL+ICL+NOISE image than in the ICL+NOISE image. This means that the superposition of ICL and galaxies diminishes the amount of flux \texttt{DAWIS} is able to detect and restore. This is expected, since \texttt{DAWIS} detects and restores first the brightest objects in the image (the galaxies). Therefore the residuals of these bright sources degrade the detection of the ICL. Additionally, the rough separation based on $j_{\rm{sep}}$ shows limitations for concentrated profiles ($r_{e}=50$~kpc). As their characteristic size is smaller, the inner part of these bright and concentrated ICL profiles is not classified as ICL but as galaxy. This leads to larger bias values than for the extended profiles.
The values of \eCgal are displayed in Figure~\ref{fig:gal_relative_error_contamination}. For the PF method, the values of \eCgal\ are always equal to zero, since the measured flux \fluxgal is equal to \fluxgalsep (see Section~\ref{subsec:perfect_fitting} for details). For \texttt{DAWIS}, the bias values are also very low. In contrast with the values of \eCICL, this shows that the superposition of galaxies and ICL has little effect on the restoration of galaxies, as explained in the previous paragraph. The method giving meaningful \eCgal values is the SBT method. The contamination effect of ICL masked with galaxies is strongly seen for bright and concentrated ICL profiles \bright\ and once again we had to truncate the bias values to fit them into the plot.

The resulting ICL fractions $f_{\rm{ICL}}$ are displayed in continuous lines in Figure~\ref{fig:ficl_all_group}, together with the ICL fractions \Ficlsep previously computed from the ICL+NOISE and GAL+NOISE images. First, we address the differences in values of \Ficlsep and \Ficl for each method. The greatest discrepancies are seen for the SBT method. As previously shown, this is mainly due to the co-addition of several contamination effects: galaxy outskirt contamination, undetected faint galaxies, ICL partially masked with galaxies for bright and concentrated profiles \bright . This gives SBT fractions \Ficl that are systematically higher than their \Ficlsep counterpart, and even higher than $f^{\rm{input}}_{\rm{ICL}}$ for low input ICL fractions ($f^{\rm{input}}_{\rm{ICL}}=0.2$). This behaviour also unveils the fact that some of the apparently accurate \Ficl values measured by the SBT method are misleading, as the detected flux associated to ICL does not come from the actual ICL profile but in vast majority from contamination effects. As they are integrated values, the measured ICL fractions in these cases are close to the input ones, but these are coincidences due to the set of input parameters. This shows once again an effect of bias compensation, where the flux belonging to ICL is not detected but is replaced by flux from galaxy outskirts.

For the PF method, \Ficl\ is very close to the \Ficlsep values at low redshift, specially for the bright and concentrated profiles \bright. It is however increasingly higher with redshift, giving even higher values than $f^{\rm{input}}_{\rm{ICL}}$ at $z=0.5$. This is caused by the faint undetected galaxy contamination effect, which is the only effect contaminating this method. In the same way as the SBT method, the apparently good \Ficl\  measured for faint and extended ICL profiles at high redshift are therefore also inflated by contamination effects. However, the PF method is the best performing method for bright and concentrated profiles ($f_{\rm{ICL}}=0.6$ and $r_{e}=50$~kpc). The measured fractions then decrease when $r_{e}$ is increasing, due to the fact that the method is still background limited and that faint and extended profiles tend much more to be hidden under it. The performance of this method is therefore strongly dependent on the surface brightness of the ICL profile, and gives a good estimation of how much the flux lost under the $3\sigma$ detection limit impacts the measurement of ICL fractions.

In the case of \texttt{DAWIS}, the atom classification effect is strongly seen for bright profiles \bright, as the values of \Ficl\ are lower than the \Ficlsep values by a factor two. Once again, the effect of the rough classification is strongly seen for these profiles, even in the measure of \Ficl. \texttt{DAWIS} performs the best for extended ICL halos, even for very faint ones. This is due to the fact that wavelet analysis, contrary to regular detections in direct space (original image) uses a spatial correlation to detect sources. The top row \texttt{DAWIS} fractions are therefore better (less contaminated, see $e^{\rm{cont}}_{\rm{ICL}}$ in Figure~\ref{fig:icl_relative_error_contamination}) than the other methods, showcasing its capacity to detect faint and extended features.

\section{Example on real astronomical data}
\label{sec:illustration_on_real_data}

\begin{figure*}
\centering
\includegraphics[width=16cm]{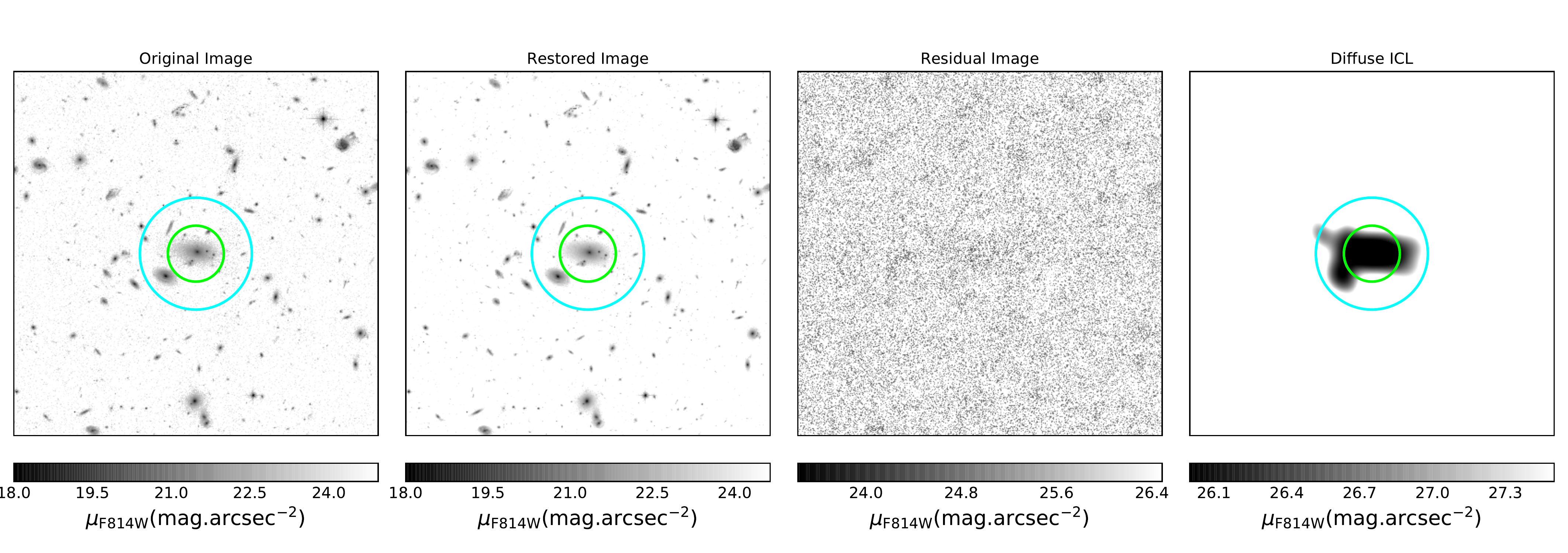}
\caption{From left to right: original astronomical image, complete restored image, residuals and restored ICL profile. The images are $1.6\times1.6$ arcmin$^2$. The green circle has a radius of 50 kpc, and the cyan circle has a radius of 100 kpc.}
\label{fig:lcdcs}
\end{figure*}

In this section we demonstrate the capabilities of \texttt{DAWIS} on a galaxy cluster image at optical wavelengths. The targeted galaxy cluster is LCDCS~0541, which is part of both the \textit{Las Campanas Distant Cluster Survey} \citep[LCDCS;][]{Gonzalez2001} and the \textit{Dark energy American French Team~/~French American DArk energy Team} \citep[DAFT/FADA;][]{Guennou2010} surveys. The redshift of this cluster is 0.542, and a deep 2x2 arcmin$^2$ HST Advanced Camera for Survey (ACS) image in the F814W filter is available. The photometric depth of the image was estimated \citep[][hereafter G12]{Guennou2012} to be around $\mu_{\rm{F814W}}\sim$~24 mag.arcsec$^{-2}$ for extended sources with \texttt{SExtractor} ($\sim$~1.25 arcsec, $1.8\sigma$ detection limit). Details about the data and the reduction procedure are available in G12.

We chose LCDCS~0541 and these data because this target was the case study of G12, where a multiscale analysis was performed to detect ICL in the images of ten galaxy clusters using \texttt{OV\_WAV}, better suited for this task than \texttt{SExtractor}. By running \texttt{DAWIS} on the same exact cluster image, we can compare the results of \texttt{DAWIS} to those obtained by this earlier multiscale analysis method (for more details on the differences between both approaches, see Section~\ref{sec:intracluster_light_detection} for traditional multiscale analysis methods, and Section~\ref{sec:dawis} for an in-depth description of \texttt{DAWIS}).

In G12, \texttt{OV\_WAV} was run two consecutive times on the LCDCS~0541 image. The first run allowed the modeling of sources with characteristic sizes up to $j=10$ (1024 pixels), and the second run up to $j=7$ (128 pixels). The detected and restored sources were removed from the original scene, and the ICL was detected by applying a 2.5$\sigma$ detection limit on the residuals. The background level for the detection limit was computed in a wide annulus in the outskirts of the residuals. The total flux in the detected ICL source was measured, and the associated absolute magnitude was computed assuming a $k$-correction of an early-type galaxy: this gave an absolute magnitude of $-20.4$. Note that with this procedure, the multiscale analysis made by the authors is analogous to a PF detection method, as the ICL sources are detected by applying a detection limit in the direct space residuals after fitting and removing all bright and small sources (MW stars and galaxies).

We run \texttt{DAWIS} on the same LCDCS~0541 F814W image, and compare both the absolute magnitude value and morphological properties of the resulting ICL source. In our case, we resize the image to a FoV of $2048\times2048$ pixel size, corresponding to $\sim1.6\times1.6$ arcmin$^2$. The \texttt{CLEAN} factor is set to $\delta=0.5$, and the relative threshold parameter is set to $\tau=0.8$. As we are only interested here in detecting the diffuse ICL source in the central part of the image, we do not apply any particular criterion for the classification step, except for a separation based on the wavelet scale of the interscale maximum $j_{\rm{sep}}$, and for the fact that the restored sources composing the atoms associated to ICL must be centered on the BCG position. We set $j_{\rm{set}}=8$ here, corresponding to a characteristic size of 256 pixels ($\sim$~80 kpc at $z=0.542$).

The resulting restored astronomical field, residual image and restored ICL distribution are displayed in Figure~\ref{fig:lcdcs}. The restored ICL distribution is anisotropic, with a boxy shape of $\sim200\times100$ kpc$^2$ and has an average apparent surface brightness of $\sim$~26.3 mag.arcsec$^{-2}$ (corresponding to a mean surface brightness of $\sim$~24.5 mag.arcsec$^{-2}$ when correcting for the cosmological dimming). The morphology of our restored ICL distribution agrees with the apparent morphology of the detection made by G12 (see their Figure~4), but ours is continuous and seems slightly more extended. We compute the associated absolute magnitude using the same $k$-correction as G12, and find a value of $-22.4$. This value is two orders of magnitude brighter than the source detected in G12. We adopt the same cosmology than G12 to compute these numbers. This detection shows the capabilities of \texttt{DAWIS} to detect these faint and extended sources where other multiscale methods could not before, and this in an automated way.

We now discuss the implications of this comparison. We detect with \texttt{DAWIS} more flux than G12, resulting in a brighter absolute magnitude. This can be attributed to two things: \texttt{DAWIS} detecting more flux due to the detection of the ICL component directly in the wavelet space, and differences in classification. First, it is expected that \texttt{DAWIS}, due to the synthesis-by-analysis and the meticulous removal of all bright sources before detecting faint ones, detects more flux in faint and extended sources than what is detected with a simple detection limit in the direct space residuals. Second, such a careful approach allows to clearly assess and classify each restored source. In the case of G12, all the sources in an \texttt{OV\_WAV} pass are restored at once, meaning that no classification is done and that signal from ICL may have been removed by this approach (especially in the first pass, that allows the removal of objects with characteristic sizes up to 1024 pixels). This also explains such a discrepancy in flux and absolute magnitude.

Of course, the question of the classification criteria, and more specifically of the separation wavelet scale $j_{\rm{sep}}$ and its value must be brought up again. We chose $j_{\rm{sep}}=8$ because a lower value would imply a significant increase in the flux associated to ICL and of its average apparent surface brightness (jumping to $\mu\sim25.7$~mag.arcsec$^{-2}$). The source we detected has already a bright average surface brightness, compared to results from literature. More importantly, the bright inner profile of the BCG of LCDCS~0541 has a spatial extent of $\sim$~40 to 50 kpc. Associating to ICL sources with characteristic sizes of at least twice this spatial extent seems to be an accurate choice. On the other hand, a value of $j_{\rm{sep}}=7$ corresponding to a typical size of $\sim$~40 kpc sets a precarious separation. In any case, the question of whether this extended LSB source should actually be called ICL or is rather the outskirts of the BCG intensity profile is still pending, and cannot really be answered with certainty with only monochromatic data.

\section{Conclusions}
\label{sec:conclusions}

We presented in Section~\ref{sec:dawis} our novel detection algorithm \texttt{DAWIS}, based on a synthesis-by-analysis approach with an operating mode based on an isotropic wavelet dictionary and on interscale connectivity analysis. The algorithm is developed keeping in mind the limitations and defects of previous wavelet algorithm packages, and therefore integrates a semi-greedy structure with parallelized modules. For an input image, \texttt{DAWIS} computes in this iterative procedure the synthesised dictionary composed of atoms (e.g. the sum of restored source profiles) and the associated denoised restored image of the original field. We showed that there is a great range of possibilities to separate objects and compose different synthesised dictionaries based on astrophysical priors and on the goals of the performed analysis.

To estimate the efficiency of the algorithm, we simulated in Section~\ref{sec:simulations} mock images of galaxy clusters with noise using the modeling package \texttt{Galsim}. Galaxies are represented by S\'ersic profiles, and the ICL by a large and faint exponential profile, the properties of which are controlled by few parameters, notably an input ICL fraction. The images are only emulating the MegaCam instrument properties, but the same study with simulations of other instruments should be led in the future. We take advantage of this work to test in Section~\ref{sec:application_of_detection_methods} the efficiency of other detection methods that we roughly classified in two categories: surface brightness threshold and fitting algorithms. We apply these two methods in the same way as \texttt{DAWIS} to the simulations, and compare the resulting measured fluxes and ICL fractions.

We showed in Section~\ref{sec:application_of_detection_methods} the limitation of the three detection methods, especially the SBT method which applies a constant threshold in surface brightness to the image to separate ICL from galaxies. As already shown in \citet[][]{Rudick2011}, galaxy light profiles and ICL are overlapping and superimposed, and the use of a single threshold to separate these two components leads to strong contamination of the detected ICL (see Section~\ref{subsec:results_superimposed}). While we managed to list at least part of the contamination and degradation effects (masks, contamination by galaxy outskirt profiles, contamination by faint undetected galaxies, loss of ICL flux under the $3\sigma$ detection limit), these effects are co-added, and future studies should be led to isolate and constrain them in a more refined way.

The PF method, acting as an upper limit to the fitting method quality, allowed us to showcase in Section~\ref{subsec:results_separated} the amount of flux lost under the detection limit when applying a standard detection procedure in the mock images. The differences on the measures of ICL fractions for faint and extended ICL profiles between this method and \texttt{DAWIS} come mainly from the fact that \texttt{DAWIS} detects sources in the wavelet scales where spatial correlation information is used. This gives food for thoughts on the importance of the methodology used to estimate the limiting depth and detection level in images, a topic already addressed by \citet{Mihos2019} and not specific to ICL, but rather touching all low surface brightness observations. We advocate here that in order to better constrain ICL detection, future studies should concentrate (in parallel with methods to separate galaxies from ICL) on the effects of sky background noise estimation in images, specially in more complex and realistic cases than the flat background of our images.

We showed that in most cases, \texttt{DAWIS} does a good job to measure ICL fractions as little contaminated as possible (see Section~\ref{subsec:results_superimposed}). However, a limitation of the priors we use to separate galaxies from ICL (based on the scale of the interscale maxima as detailed in Section~\ref{subsec:dawis_tuned}) becomes apparent when \texttt{DAWIS} tries to process images with bright and concentrated ICL halos; part of the synthesis atoms that should be associated to ICL are in this case wrongly classified and associated to galaxies. The classification operator used in this work is simplistic in nature and based mainly on the characteristic size and spatial information of the processed atoms. This fact is accentuated by our results on LCDCS~0541, for which we discussed the relevance of such a prior to separate ICL from the BCG intensity distribution in real astronomical data. The large discrepancy in absolute magnitude between our detection and the detection made by G12 could be in part explained by the different classifications made in both studies. In future studies, we plan to implement a more complete and robust classification for the ICL, using additional criteria such as morphology or granularity for example.

As all methods show flaws in the detection of ICL, we want to emphasize here the notion of bias compensation.  Some ICL fractions measured with the PF or the SBT method appear accurate (see Section~\ref{subsec:results_superimposed}) not because the measured ICL and galaxy fluxes are correct, but because different contamination and degradation effects are compensating each other: because ICL fractions are fractions of integrated values, an effect of bias compensation is occurring. This effect of bias compensation cannot really be seen or estimated in real astronomical images, which is problematic. This would definitely have worrisome effects in multi-band studies (ICL spectral energy distribution for example), where evolution of ICL fractions with wavelength could be misinterpreted. Our conclusion here is that the ICL fraction taken individually, is not a good metric  to characterise ICL. We advocate for the use of physical flux values in future studies for both galaxies and ICL, in order to better put in evidence the results of different approaches to detect ICL.

Our simulations in this work do not integrate several other components or contamination effects in the detection of LSB sources: large-scale background spatial variation, bright foreground stars, scattered light halos, presence of tidal streams just to cite a few. While the isotropic wavelet dictionary is a good tool to detect and restore quasi-circular sources such as galaxies, these other features are not particularly well expressed in the associated representation space. However, the core operating mode of \texttt{DAWIS} is versatile enough to allow future upgrades for these components, not only on the atom classification step to limit contamination effects, but also in the analysis dictionaries used for the detection or the restoration of sources. Including analyses based on other representation spaces and function bases adequate for these signal shapes is also a good direction to take when trying to capture the low surface brightness content of astronomical images.

\begin{acknowledgements}
We acknowledge long-term support from CNES. N.~Martinet acknowledges support from a CNES fellowship. This work has made use of the CANDIDE Cluster at the Institut d'Astrophysique de Paris and was made possible by grants from the PNCG and the DIM-ACAV. 
\end{acknowledgements}


\bibliographystyle{aa}  
\bibliography{aa} 

\appendix

\section{Varying SBT}
\label{sec:varying_sbt}
\begin{figure*}
  \centering
  \includegraphics[width=16cm]{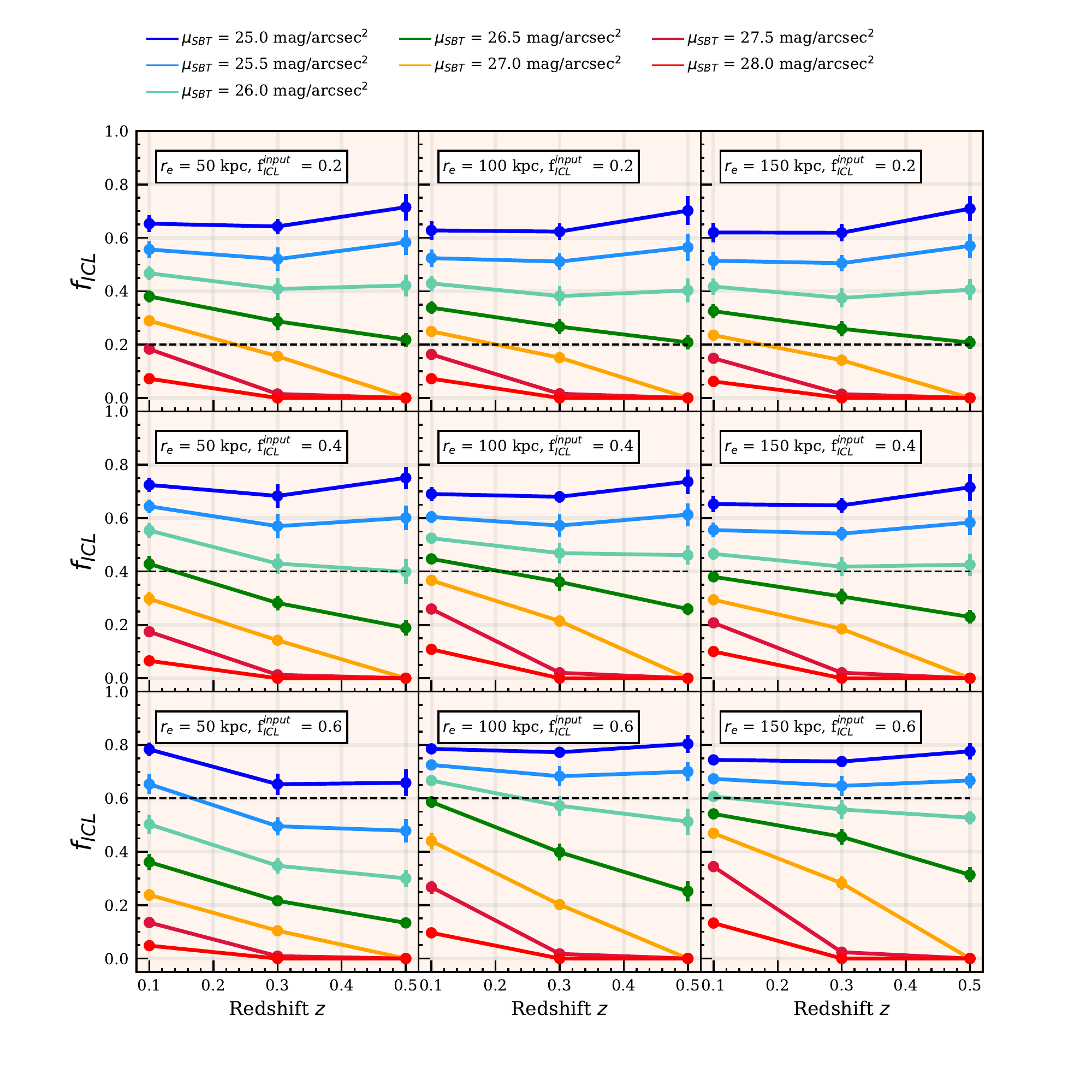}
  \caption{ICL fractions displayed for the different ICL profile parameters (half-light radius $r_{e}$, input ICL fraction $f^{\rm{input}}_{\rm{ICL}}$ and redshift $z$) as measured with different SBT. The circles with continuous lines are the ICL fractions \Ficl computed from each of the 270 GAL+ICL+NOISE MegaCam-type images, and the displayed values correspond to the average of the ten images with the same input parameters. The error bars correspond to standard deviations across ten different clusters. The input ICL fraction is also displayed as a black horizontal dashed line for better visualization.}
  \label{fig:ficl_all_sbt}
\end{figure*}

Different values have been tested for the SBT method with $25 \le \mu_{\rm{SBT}} \le 28$. We applied exactly the same methodology as in Section~\ref{subsec:SBT}, and measured the ICL fractions in the GAL+ICL+NOISE images. The results are shown in Figure~\ref{fig:ficl_all_sbt}, with the same format as Figure~\ref{fig:ficl_all_group}. As expected and consistent with the literature \citep{Rudick2009, Tang2018}, the choice of SBT influences greatly the measured ICL fraction, with values ranging from $\sim$~0\% to $\sim$~80\% in some cases for the same input ICL fraction. Most of the times however, thresholds \sbt~ or $\mu_{\rm{SBT}}=27$~mag.arcsec$^{-2}$, which are the generic SBTs considered in the literature, give results that are more consistent with the input ICL fraction. However, this trend needs to be contrasted with our in depth analysis of biases in the measure of \fluxgal~ and \fluxicl~ (see Section~\ref{subsec:results_separated} and Section~\ref{subsec:results_superimposed}).

\end{document}